\documentstyle[12pt,aaspp4]{article}
\begin{document}

\title{\bf Multiwavelength Observations of the Second Largest Known
FR II Radio Galaxy, NVSS 2146+82}
\author{Christopher Palma\altaffilmark{1}, Franz E. Bauer\altaffilmark{2}}
\affil{Department of Astronomy, University of Virginia}
\authoraddr{P. O. Box 3818, Charlottesville, VA  22903-0818}
\altaffiltext{1}{NOAO WIYN Queue Investigator}
\altaffiltext{2}{National Radio Astronomy Observatory Jansky
Pre-Doctoral Fellow} 

\author{William D. Cotton,  Alan H. Bridle}
\affil{National Radio Astronomy Observatory\altaffilmark{3}}
\authoraddr{520 Edgemont Road, Charlottesville, VA 22903-2475}
\altaffiltext{3}{The National Radio Astronomy Observatory is a facility
of the National Science Foundation operated under cooperative agreement
by Associated Universities, Inc.}

\author{Steven R. Majewski\altaffilmark{4}, \& Craig L. Sarazin}
\authoraddr{P. O. Box 3818, Charlottesville, VA  22903-0818}
\affil{Department of Astronomy, University of Virginia}
\altaffiltext{4}{David and Lucile Packard Foundation Fellow; Cottrell
Scholar of the Research Corporation; National Science Foundation
CAREER Fellow; Visiting Associate, The Observatories of the Carnegie
Institution of Washington}

\begin{abstract} 

We present multi-frequency VLA, multicolor CCD imaging, optical
spectroscopy, and {\it ROSAT} HRI observations of the giant FR II radio
galaxy NVSS 2146+82.  This galaxy, which was discovered by the NRAO VLA
Sky Survey (NVSS), has an angular extent of nearly 20\arcmin\ from lobe
to lobe.  The radio structure is normal for an FR~II source except for
its large size and regions in the lobes with unusually flat radio
spectra.  Our spectroscopy indicates that the optical counterpart of
the radio core is at a redshift of $z=0.145$, so the linear size of the
radio structure is $\sim$4$h_{50}^{-1}$ Mpc, $H_{0} = 50h_{50}$.
This object is therefore the second largest FR II known (3C 236 is
$\sim$6$h_{50}^{-1}$ Mpc).  Optical imaging of the field surrounding
the host galaxy reveals an excess number of candidate galaxy cluster
members above the number typically found in the field surrounding a
giant radio galaxy.  WIYN HYDRA spectra of a sample of the candidate
cluster members reveal that six share the same redshift as NVSS
2146+82, indicating the presence of at least a ``rich group''
containing the FR II host galaxy.  {\it ROSAT} HRI observations of NVSS
2146+82 place upper limits on the X-ray flux of
$1.33\times10^{-13}$ ergs cm$^{-2}$ s$^{-1}$ for any hot IGM and
$3.52\times10^{-14}$ ergs cm$^{-2}$ s$^{-1}$ for an X-ray
AGN, thereby limiting any X-ray emission at the distance of the radio
galaxy to that typical of a poor group or weak AGN.  Several other
giant radio galaxies have been found in regions with overdensities of
nearby galaxies, and a separate study has shown that groups containing
FR~IIs are underluminous in X-rays compared to groups without radio
sources.  We speculate that the presence of the host galaxy in an
optically rich group of galaxies that is underluminous in X-rays may be
related to the giant radio galaxy phenomenon.

\end{abstract}


\section{\bf Introduction}

The ``giant'' radio galaxies (GRGs), which we define as double radio sources
whose overall projected linear extents exceed 2$h_{50}^{-1}$ Mpc, are 
interesting as extreme examples of radio source development and evolution.
Members of this class, which comprise only a few percent of all
powerful extragalactic radio sources, have been documented for almost
25 years (e.g., \cite{wil74}).  They have been used to constrain the
spectral aging and evolution of radio sources and as tests for the
evolution of conditions in intergalactic environments on Mpc scales
(\cite{str80}; \cite{sub93}; \cite{cot96}).  Their 1.4 GHz radio powers
are generally in the regime $10^{24.5}< P_{1.4}<10^{26}$ $h_{50}^{-2}$ W Hz$^{-1}$,
just above the transition between Fanaroff-Riley Types I (plumed) and
II (lobed) radio structures (\cite{fan74}).  It is unclear whether the
giant sources are examples of unusually long-lived (and directionally
stable) nuclear activity in radio-loud systems, or of the development
of sources in unusually low-density environments.

Because of their large angular sizes, nearby giant radio galaxies can
in principle be studied in great detail, but their
largest-scale structures may be over-resolved and undersampled by
interferometers.  They have traditionally been discovered through sky
surveys with compact interferometers or single dishes at relatively
low frequencies, where angular resolution is modest but large fields
of view and diffuse steep-spectrum structures can be imaged more
easily.  The source NVSS 2146+82 was noted as a candidate giant radio
galaxy when it appeared in the first 4$^\circ$\ by 4$^\circ$\ field
surveyed by the NRAO VLA Sky Survey (NVSS: \cite{Condon98}), a
northern-hemisphere survey at 1.4 GHz using the VLA D configuration at
45\arcsec\ (FWHM) resolution.

\begin{figure}
        \plotfiddle{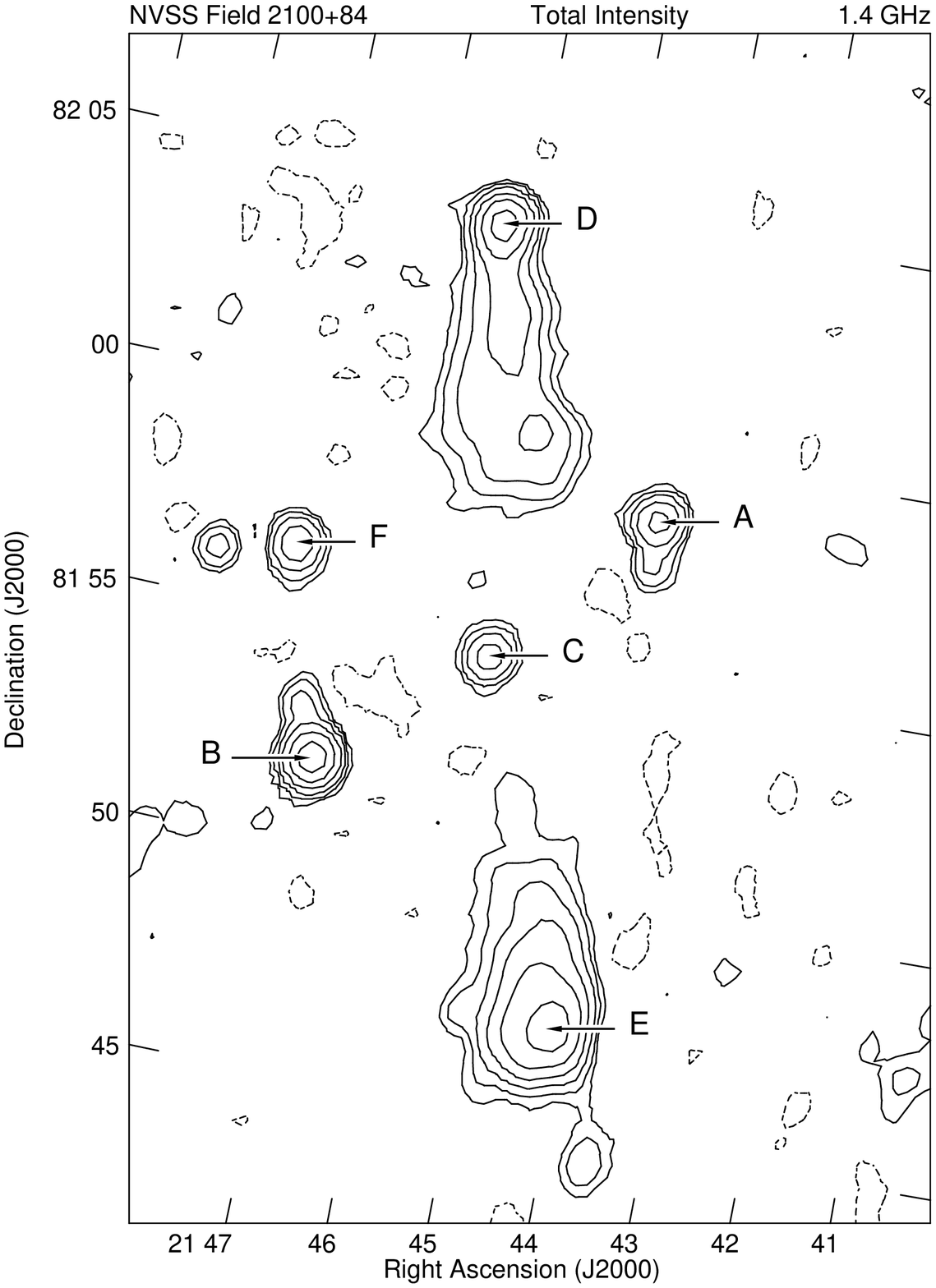}{6truein}{0}{70}{70}{-220}{-50}
	\caption{Contour plot of the NVSS 1.4 GHz 
	total intensity data for the field. Contours are
	shown at -1, 1, 2,4, 8, 16, and 32 mJy per CLEAN beam
	area.}
 \label{fig:NVSScontours}
 \end{figure}

Figure \ref{fig:NVSScontours} shows contours of the NVSS image at
45\arcsec\ resolution.  There are two symmetric, extended lobes (D and
E) on either side of an unresolved component C, plus an unusually
large number of other radio sources within 10\arcmin\ of C.  Two of
these (A and B) are also symmetrically located around C.

Comparison with the Digital Sky Survey (DSS) showed that source C
coincides with an $\sim$18$^{\rm th}$ mag elliptical galaxy to within
the uncertainties in the NVSS and DSS positions.  If the elliptical
galaxy is the host of an unusually large radio source (C+D+E), then
the apparent magnitude suggests that the whole structure may be
similar in linear scale to 3C\,236.  The DSS also shows a nearby image
that might be another galactic nucleus, and a faint extended feature
suggesting a possible ``tail'' or interaction.

We have undertaken several observational studies of the radio and
optical objects in the field to determine their nature and to clarify
the relationships between the optical and radio sources.  These studies
include:

\begin{enumerate}
\item High resolution radio imaging at 4.9 and 8.4 GHz to locate any 
compact flat-spectrum radio components in the field, and thus to 
identify any AGN that could be responsible for some or all of the 
other radio emission,
\item A search for fainter diffuse radio emission between the D and E
components that might link them together or to other sources in the
field and thus clarify their physical relationship,
\item Higher-resolution radio imaging of the other radio sources in the
field to explore whether they might be physically related to the
diffuse components, or to each other by gravitational lensing,
\item Optical spectroscopy of both optical ``nuclei'' and other galaxies in
the field,
\item UBVRI optical photometry of the field, and
\item X-ray imaging using {\it ROSAT} HRI observations to search for
any hot X-ray emitting gas which might be associated with an overdensity of
galaxies or non-thermal X-ray emission from an AGN.
\end{enumerate}

Throughout this paper, we assume a Hubble constant $H_{0} = 50h_{50}$ km
s$^{-1}$ Mpc$^{-1}$ .
At a redshift of $z = 0.145$, the angular diameter
distance to the radio galaxy is 708.4$h_{50}^{-1}$ Mpc, the luminosity distance is
928.7$h_{50}^{-1}$ Mpc, and $1\arcmin$ corresponds to 206$h_{50}^{-1}$ kpc.

\section{\bf Radio Observations}
Table \ref{VLAObsLog} gives a journal of our VLA observations.  The
observations in the A configuration were designed to locate any 
compact radio components in the field. Those in the B, C, and D
configurations were intended to image the largest scale emission in
enough detail to reveal any relationships and connections between the
extended components, as well as to determine their spectral and
Faraday rotation/depolarization properties.  The BnC configuration
data were designed as a sensitive search for connections, such as
jets, between the central radio source and the extended features.

The flux density calibration was based on 3C\,48 and 3C\,286.  The
on-axis instrumental polarization corrections were determined from
observations of the unresolved synthesis phase calibrator 2005+778,
and the absolute polarization position angle scale from observations
of 3C\,286.  Multiple observations of 3C\,286 and other polarized
sources were used to detect problems with ionospheric Faraday
rotation, but none was noted in any of the sessions.  
The data were calibrated using the source 2005+778 as an intermediate
phase reference, then self-calibrated using AIPS software developed by
W. D. Cotton for the NVSS survey.

Due to the large size of this source, 1.4 and 1.6 GHz observations
used three pointings; one on the central source C, and one near the
center of each putative lobe.  The B, C, and D VLA configuration
observations were made at 1.365 and 1.636 GHz to measure rotation
measure and spectral index.  The data from these frequencies were
calibrated and imaged separately.  Data taken in the BnC configuration
were in two adjacent 50 MHz bands centered on 1.4 GHz.  Since the
source extent is comparable to that of the antenna pattern and the
bandwidth used was relatively large, the deconvolution (CLEAN) and
self calibration applied corrections for the frequency dependence of
the antenna pattern.  Data from each of the three pointings were
imaged independently and combined into a single image by interpolating
the images onto a common grid, averaging weightings by the square of the
antenna power pattern, and correcting for the effects of the antenna
pattern.  The 0.3 GHz observations were of limited use owing to
interference.

\subsection{Radio Results}
The most sensitive image of NVSS 2146+82 is derived from our BnC
configuration data at 1.4 GHz which has a resolution of 13\arcsec\
(FWHM).  Figure \ref{fig:NVSSLhi} shows logarithmic contours of the
total intensity in the region around the source in this image; the rms
noise is 20 $\mu$Jy per CLEAN beam area.  A gray scale representation
of the same image showing the filamentary structure of the lobes is
given in Figure \ref{fig:BnCgray}.  Figure \ref{fig:NVSSjets} shows
the inner region of this image contoured to lower levels using an
initially linear contour interval.

\subsection {Association of Features}
\begin{figure}
        \plotfiddle{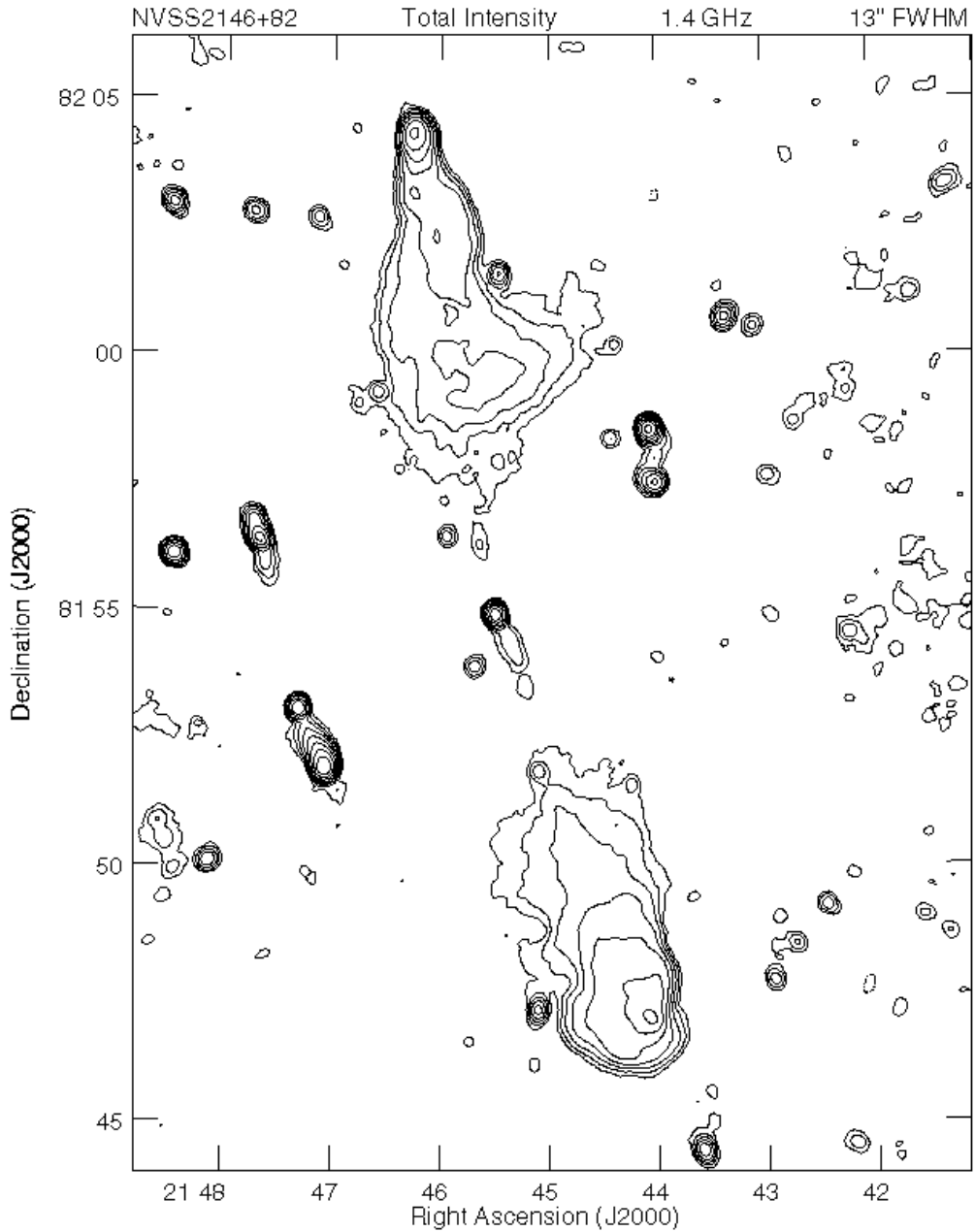}{6truein}{0}{70}{70}{-220}{-50}
	\caption{Contour plot of the new 1.4 GHz 
	total intensity data for the field at 13\arcsec\ (FWHM)
	resolution. Contours are
	shown at -1, 1, 2,4, 8, 16, 32 and 64 times 100 $\mu$Jy per CLEAN beam
	area.}
 \label{fig:NVSSLhi}
 \end{figure}

\begin{figure}
        \plotfiddle{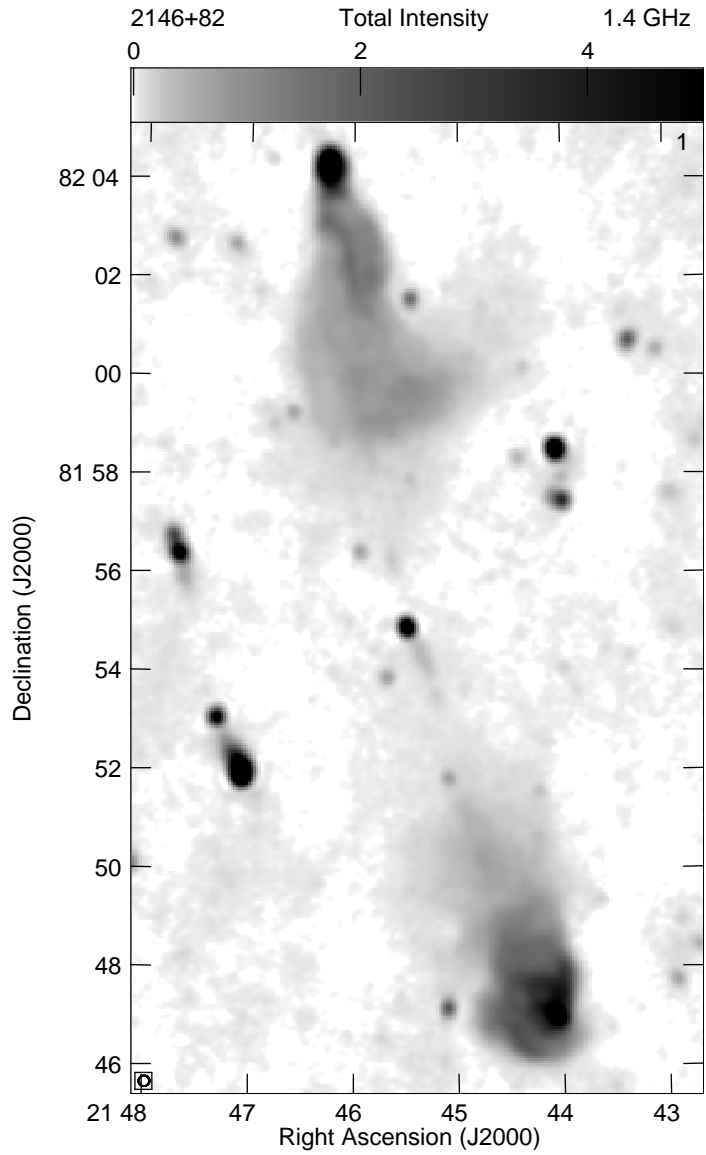}{6truein}{0}{70}{70}{-220}{-50}
	\caption{Gray scale image at 13\arcsec\ (FWHM) resolution using
        a nonlinear transfer function to emphasize the lower
        brightness levels.
        The jet and strong filaments in the lobes can be seen.
	}
 \label{fig:BnCgray}
 \end{figure}

The structures of the extended features D and E shown in Figures
\ref{fig:NVSSLhi} and \ref{fig:BnCgray} are entirely consistent with
their being associated with each other as the two lobes of a large
FR\,II double source of overall angular size 19\farcm5.  Both features
are brightest in the regions furthest from C, contain bright (but
resolved) substructure near their outer edges resembling the hot spots
of FR\,II sources, and have their steepest brightness gradients on
their outer edges.  The overall length of the two lobes is the same to
within 5\%.  Although features A and B in Figure \ref{fig:NVSScontours}
appear symmetric around feature C, the higher resolution VLA images 
(Figures \ref{fig:NVSSLhi} and \ref{fig:BnCgray}) reveal
them to be background sources, unrelated to NVSS 2146+82.

The northern feature (D) contains a region of enhanced emission (hot
spot) at its northern extremity with about 65 mJy in an area 30\arcsec\
by 18\arcsec\, and an L-shaped extension to the West.  The southern
feature (E) has 75 mJy in a region of enhanced emission 50\arcsec\ by
30\arcsec\ (a ``warm spot'') recessed by 10\% of the distance from the core 
and sharp brightness gradients around its southern and
western boundaries.  Both regions of enhanced emission show evidence of finer, but
resolved, structure in our data taken in the B configuration (see
contour plots in Figure \ref{fig:hotspots}). Figure \ref{fig:BnCgray}
clearly shows that the internal brightness distributions of both lobes
are non-uniform, and suggest the presence of filamentary structures,
again a common characteristic of FR\,II radio lobes at this relative
resolution.

\begin{figure}
        \plotfiddle{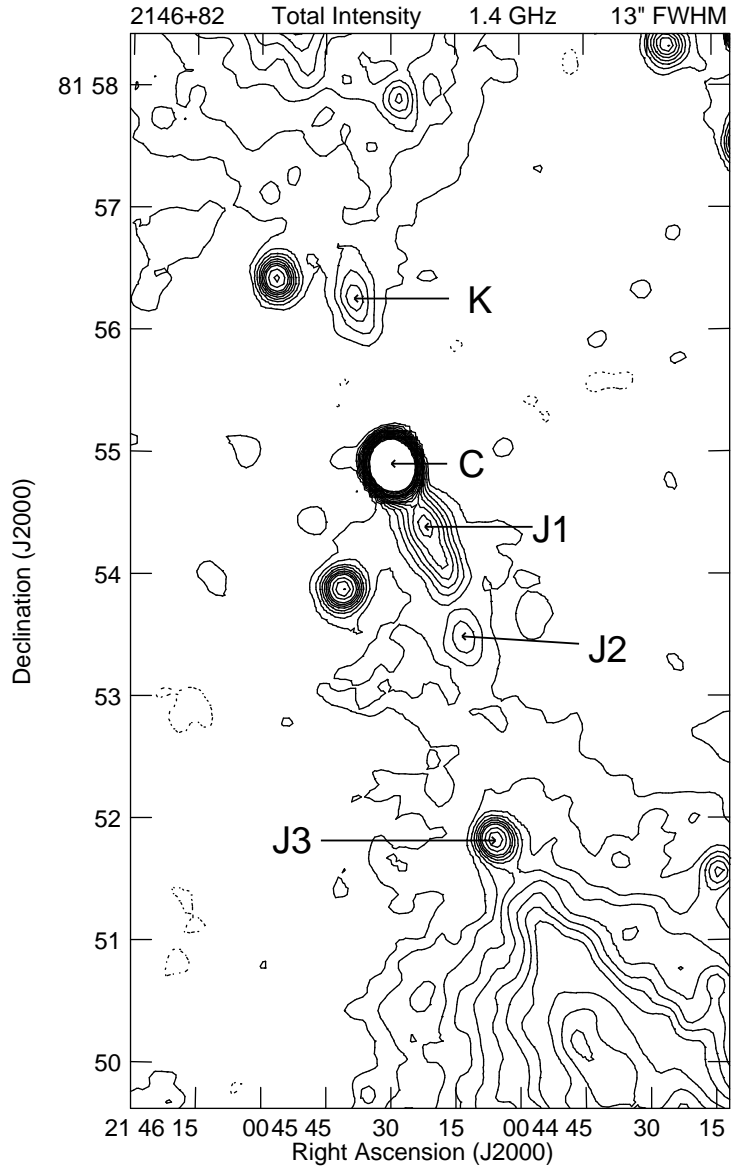}{6truein}{0}{70}{70}{-220}{-50}
 	\caption{Contour plot of the new 1.4 GHz 
	total intensity data for the field at 13\arcsec\ (FWHM)
	resolution. 
        Contours are shown at 
        -1, 1, 2, 3, 4, 5, 6, 7, 8, 10, and 12 times 50 $\mu$Jy per CLEAN beam
 	area.
        The core and various features in the jet are marked.}
	 \label{fig:NVSSjets}
 \end{figure}

\begin{figure}
\plottwo{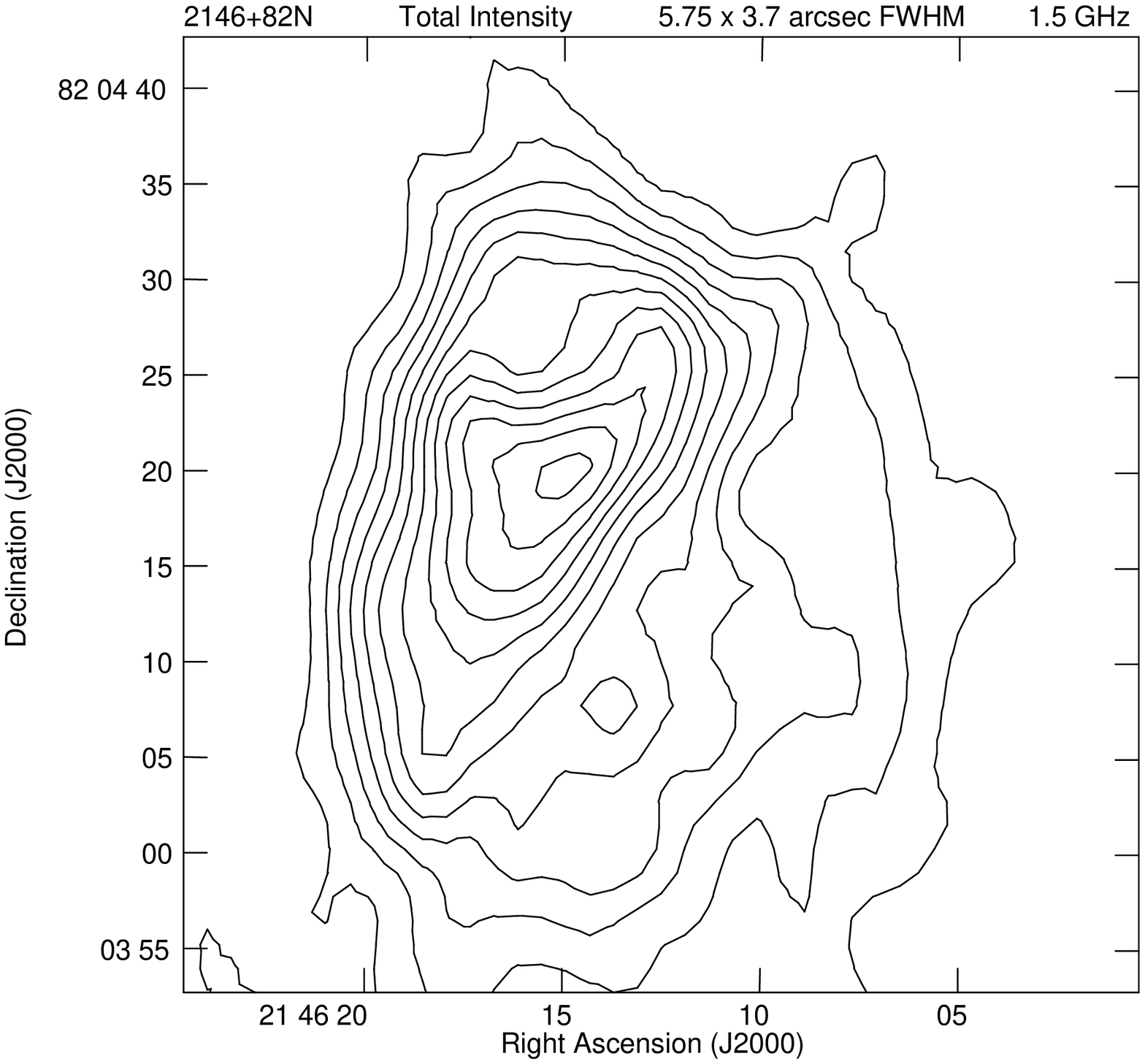}{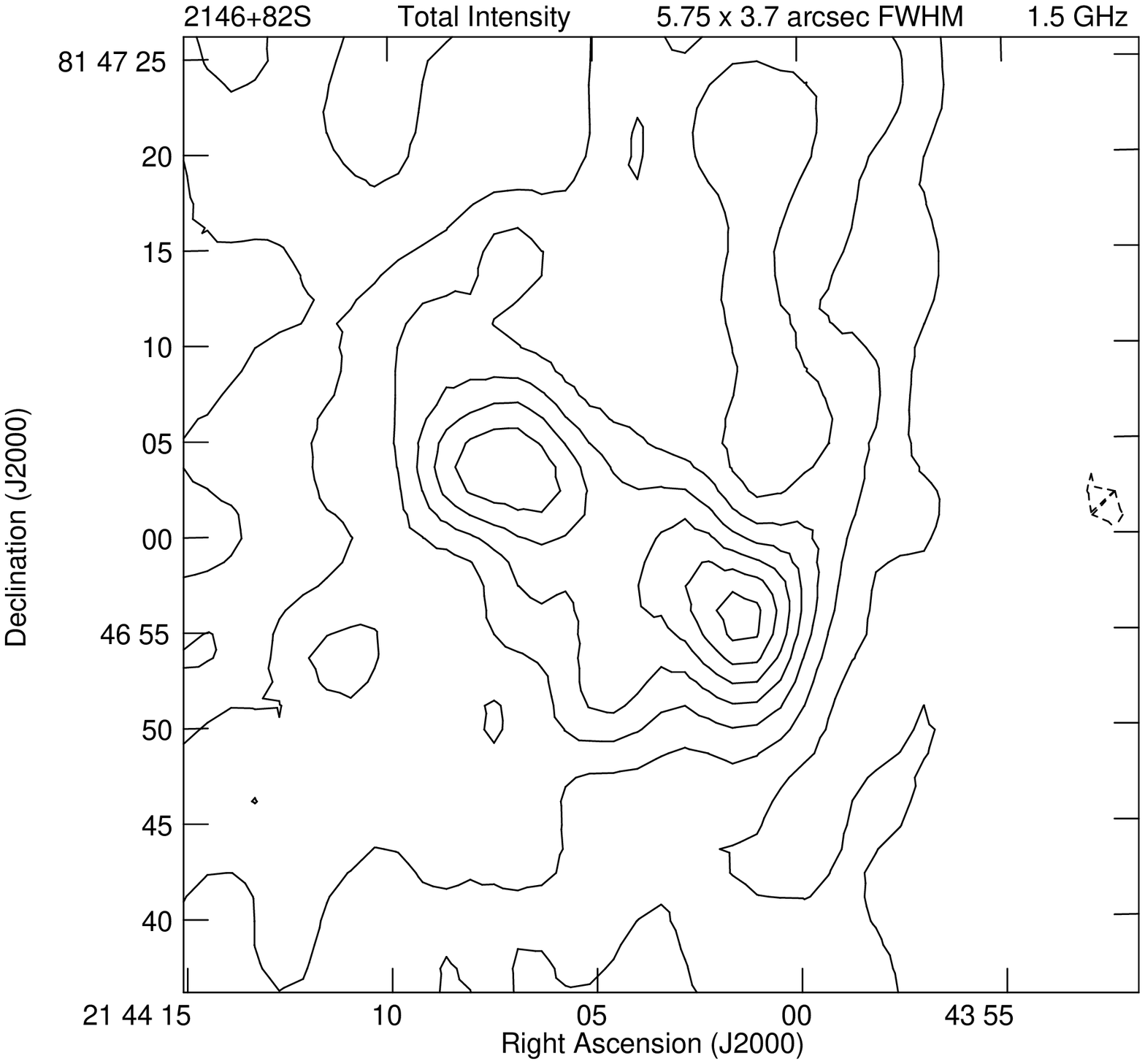}
\caption{Contour plot of the 1.5 GHz total intensity data from the B configuration 
over the north D (left) and south E (right) hot spots
of the source at 5\farcs75 by 3\farcs7
(FWHM) resolution. 
Contours are shown at a linear interval of 0.25 mJy per CLEAN beam area}
\label{fig:hotspots}
\end{figure}

Most importantly, Figures \ref{fig:NVSSLhi}, \ref{fig:BnCgray}, and
\ref{fig:NVSSjets} also show that these lobes are linked to the
central compact feature C by elongated  features that are plausibly
the brightest segments of a weak jet-counterjet system. 
These features are labeled in Figure \ref{fig:NVSSjets}.

We interpret the following features as belonging to the jet in the
south lobe.

{\bf J1}. This feature is clearly part of a jet that points towards
the south lobe but not directly at the peak of feature E.

{\bf J2}. 
This feature (1\farcm5 from C) and feature K (1\farcm4 to the
north of C) are roughly symmetric in distance from C and in intensity
but are not quite collinear with C.
On both sides of the source the jet becomes harder to trace further
into the lobe.
J2 appears to be south of the C--J1 direction, suggesting a southward
bend, however.

{\bf J3}. This feature is plausibly a knot in the continuation of the
jet into the south lobe.
The lobe brightens beyond J3 and contains a diffuse ridge that is a
plausible continuation of the (possibly decollimated) jet in the
direction of the ``warm spot'' E.
The north lobe also brightens at about the same distance from C
although there is no feature  corresponding to J3 in the north.

Table \ref{table:fluxes} gives flux density estimates for the main
features of the source.  
We estimate that the jet and counterjet together comprise about 1\% of
the total flux density of the extended lobes, a typical jet
``prominence" for radio galaxies slightly above the FR\,I--II
transition. 

The higher-resolution radio images provide no evidence that sources A,
B, or F in Figure 1 are physically related to each other, or to C, D
and E.   Although none can be optically identified, we consider it
likely that these are three (or more) unrelated background sources.
The symmetrical alignment of A and B around C is apparently
coincidental, and there is no evidence for any radio ``bridge'' between
these sources and component C.

\subsection{Polarimetry}
\begin{figure}
\plottwo{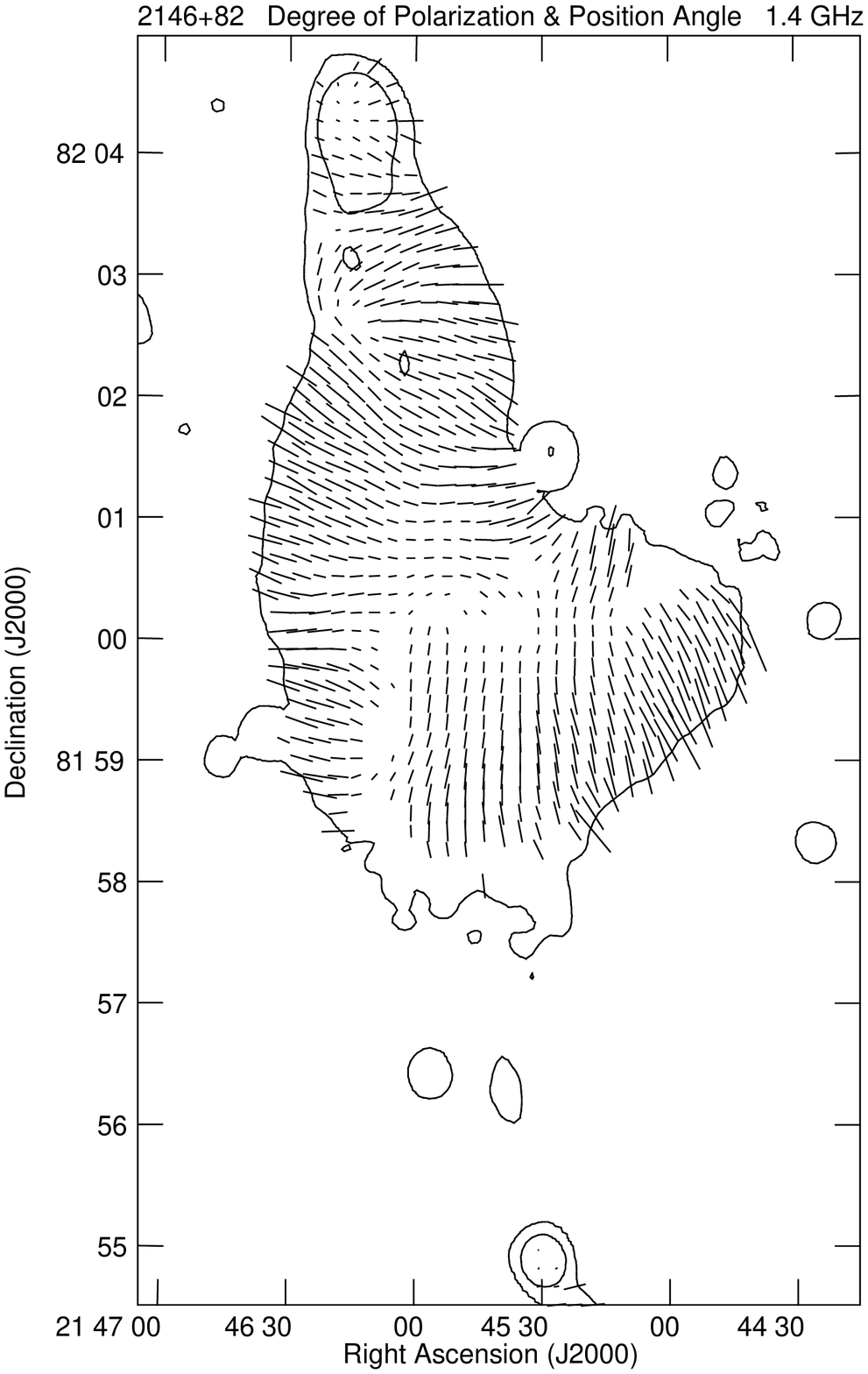}{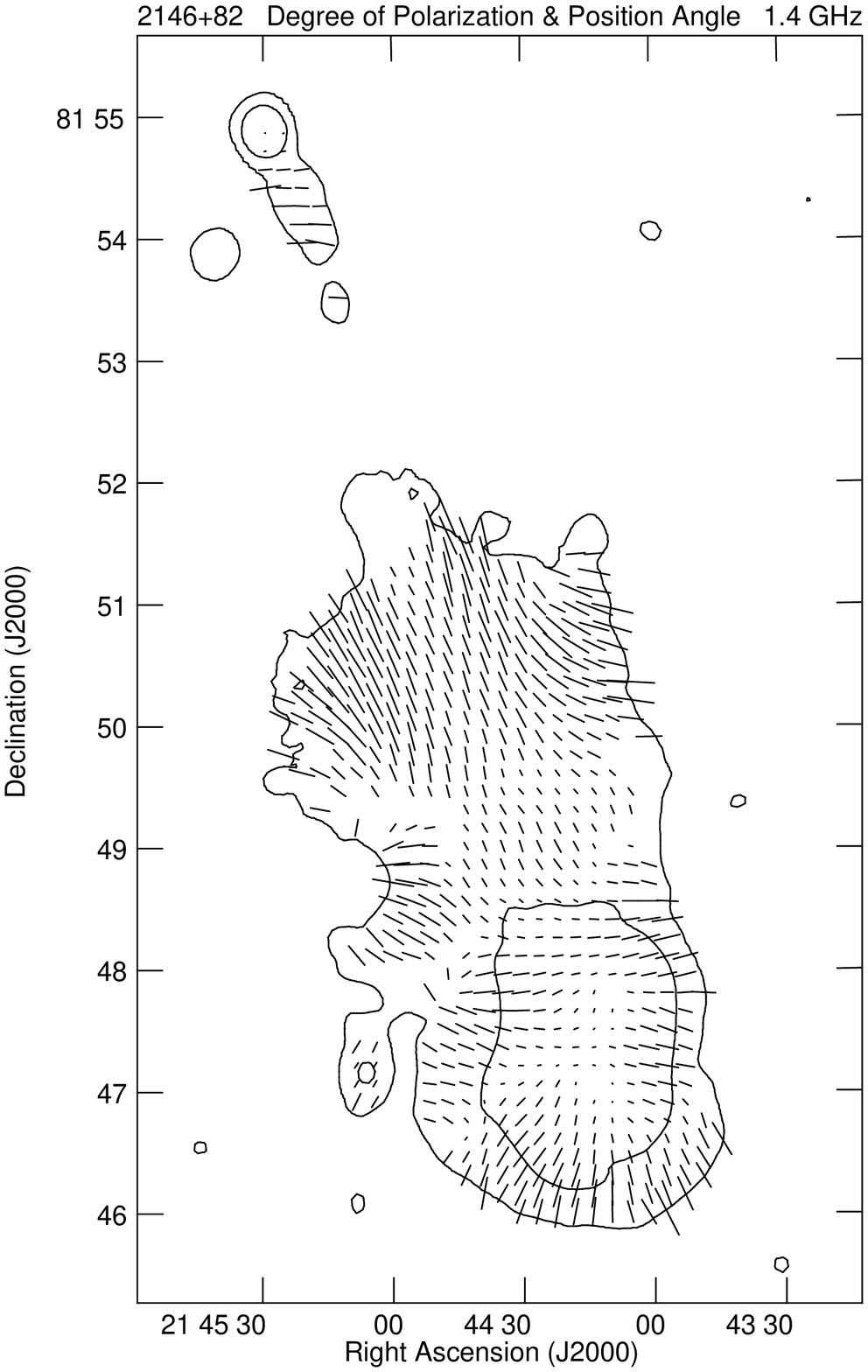}
\caption{Distribution of degree of  1.4 GHz linear polarization $p$ and {\bf E}-vector
position angle $\chi$ over the north D (left) and south E (right) 
lobes of the source at 13\arcsec\
(FWHM) resolution, superposed on selected contours of total intensity.
A vector of length 15\arcsec\ corresponds to $p$=0.5.}
\label{fig:polarization}
\end{figure}

The polarization structure derived from the sensitive BnC configuration observations
is shown in Figure \ref{fig:polarization}.  The 1.4 and 1.6 GHz data
are sufficiently separated in frequency to enable us to measure
Faraday rotation but still maintain comparable surface brightness
sensitivity.  The derived rotation measure images of the two lobes are
shown in Figure \ref{fig:RM}.  The rotation measure distribution over
the north lobe is featureless but several filamentary rotation measure
structures can be seen over the southern lobe.  The average rotation
measure is about the same in the two lobes, -9 rad\ m$^{-2}$ in the
north and -8 rad\ m$^{-2}$ in the south.  The Faraday rotation measure
in the south lobe has a somewhat larger root mean square variation, 8
rad\ m$^{-2}$ compared to 5 rad\ m$^{-2}$ in the north.

\subsection{Spectral Index Distribution}

Figure \ref{fig:WENSS} shows the 0.35 to 1.4 GHz spectral index
distribution inferred from comparing the WENSS (\cite{WENSS}) image
with our BnC configuration image convolved to the same resolution.  The northern and
southern warm spots have spectral indices\footnote{Spectral index,
$\alpha$, as used here is given by $S = S_0\nu^{\alpha}$.}
$\alpha^{1.4}_{0.35}$ of -0.6 and -0.55, not unusual for the hot spots
of FR~II sources in this frequency regime.  The background sources also
exhibit spectral indices that are quite typical of extragalactic
sources (A, -0.68; B, -1.0; F, -0.7).  Near the centers of the north
and south lobes of NVSS 2146+82, however, this comparison shows regions
of unusually ``flat'' spectral index ($\alpha^{1.4}_{0.35}\ \approx\
-0.3\ \pm 0.02$ in the north lobe, $\alpha^{1.4}_{0.35}\ \approx\
-0.4\ \pm 0.03$ in the south lobe).

The spectral index variations across the lobes can also be studied from
our 1.36 and 1.63 GHz data.  Due to the low surface brightness the data
were tapered to 55$\arcsec$ resolution before imaging for this
comparison.  To eliminate any complication from the mosaicing
technique, only data derived from the pointing on a given lobe were
used to determine the spectral index variations for that lobe.  Thus,
the data from two pointings were imaged independently at 1.36 and 1.63
GHz, corrected for the antenna power pattern, and spectral index images
were derived independently for the two lobes.  These results are shown
in Figure \ref{fig:SI}.  The close spacing of the frequencies makes
determining the spectral index more difficult; but this is compensated
to some extent by the nearly identical imaging properties at the two
frequencies, which reduce systematic errors.  These data sets are fully
independent of those used for the spectral index image in Figure
\ref{fig:WENSS}, but also reveal symmetric regions of unusually flat 
spectral index, $\alpha^{1.6}_{1.4}\ \approx$ -0.3$ \pm 0.08$, in both
lobes.

We conclude that four independent data sets show evidence for regions
with $\alpha^{1.4}_{0.35}\ \approx$ -0.3 in regions of relatively high
signal to noise ratio.  These regions are not artifacts of
``lumpiness'' in the zero levels of the images.

\subsection {Source Alignment}
NVSS 2146+82 is not aligned along a single axis.  The two warm regions
(E and D) and the core (C) are not collinear.  The jet in the south
appears to have several bends; one near the end of J1 (see Figure
\ref{fig:NVSSjets}) where it bends toward J2, a change in position
angle from -150$^\circ$ to -170$^\circ$.  Beyond J3, the ridge line of
the lobe is fairly well defined and is again at position angle
-150$^\circ$, consistent with a second bend (apparently $\approx$
20$^\circ$) in the neighborhood of J3.  The jet is not so prominent in
the north but feature K, which may be the brightest part of a
counterjet, is elongated along position angle of -169$^\circ$.

The general ``C'' shape of the source suggests that the overall
misalignment is due to environmental effects that have bent the jets,
rather than to a changing initial jet direction which is likely to
produce overall ``S'' symmetry.

We consider it beyond doubt that C, D and E comprise a single large FR\,II
radio source with weak radio jets, whose parent object is the galaxy
identified with C.

\section{\bf Optical Observations of NVSS 2146+82 and its Environs} 

Optical photometric and spectroscopic observations were obtained to
identify the host galaxy of the radio emission and to measure its
redshift.  We began the search for the optical counterpart to the radio
source using the Digitized Sky Survey (\cite{dss}; hereafter
DSS). The radio core is aligned with an elliptical galaxy on the DSS
image to within the astrometric accuracy of
the radio and optical positions from the NVSS and DSS.  There is also a
second, equally bright object a few arcseconds east of the galaxy at
the radio core position.  Finally, in the DSS image, there appears to
be S--shaped diffuse emission that passes through both bright
``nuclei''.
Therefore, our initial assumption was that
the host galaxy of NVSS 2146+82 was possibly a disturbed, double
nucleus galaxy.  In the following sections, we summarize the optical
imaging of the field surrounding the candidate host galaxy and the
spectroscopic observations of this host galaxy and its candidate
galactic companions.

\subsection{Photometric Observations} 

U, B, V, R, and I CCD observations were obtained at the 1.52-m
telescope at Palomar Observatory on the nights of 7-9 January 1997.  In
addition, U, B, V, and I CCD observations were made at Kitt Peak
National Observatory on 4 April 1997.  The Palomar 1.52-m observations
were made with a 2048 $\times$ 2048 CCD with a pixel scale of 0\farcs37
per pixel, resulting in a 12\farcm63 field of view.  Though
photometric, the seeing was poor ($2-5\arcsec\:$ on 7 January,
$1.5-2.5\arcsec\:$ on 8,9 January) during the Palomar run, so higher
resolution ($1.2-1.4\arcsec\:$ seeing) images were obtained with the KPNO
4-m telescope in April.  The KPNO observations were made with the prime
focus T2KB CCD with a pixel scale of $0\farcs47$ per pixel, resulting
in a 16$^{'}$ field of view.  Because the KPNO data were not taken in
photometric conditions, the Palomar data remained useful for
calibration.  Data from both observing runs were reduced using the 
standard IRAF CCDRED reduction tasks.

After the initial reduction, aperture photometry was performed on the
host galaxy of NVSS 2146+82 using the IRAF package APPHOT.
Unfortunately, due to the poor seeing on the first night of the Palomar
run and the proximity of the foreground star (see \S 3.3) to the AGN
host, it was impossible to photometer NVSS 2146+82 without significant
flux contamination from the foreground star.  Therefore, we used the
DAOPHOT II package (\cite{stetson87}) to PSF fit and subtract stars from
the Palomar NVSS 2146+82 images.

After the foreground star was subtracted, photometry of the galaxy
was performed identically to the photometry of several \cite{landolt}
standard stars.  Approximately 20 stars were selected from each frame
containing the AGN host galaxy.  A circular aperture 2.5 times the
average FWHM of these stars was used to measure the flux of the host
galaxy.  This aperture was chosen to be consistent with the standard
star photometry and because it completely enclosed the host without
including contaminating flux from other nearby objects.

Once instrumental magnitudes for the galaxy were determined, they were
transformed to the standard system using transformation equations
incorporating an airmass and color term that were 
determined for the Landolt standard stars.  The results of our
U,B,V,R,\& I photometry of the host galaxy are listed in Table
\ref{table:magnitudes}.

\subsection{Spectroscopic Observations}

Optical spectra of NVSS 2146+82 were obtained at Kitt Peak National
Observatory on 9 December 1996.  The spectroscopic observations were
made with the RC Spectrograph on the KPNO Mayall 4-meter telescope.
The detector in use was the T2KB CCD in a 700 $\times$ 2048 pixel
format.  All exposures were made with a $1\arcsec$ slit width and a 527
lines/mm grating.  The spectral resolution, measured using unresolved
night sky lines, is $\sim$3.4 \AA.  The data were reduced using the
standard IRAF reduction tasks.  The extracted spectra were wavelength
calibrated using a solution determined from the spectrum of a HeNeAr
comparison source.  Finally, spectrophotometric calibration was applied
using a flux scale extrapolated from several standard star spectra.

Spectra of candidate galactic companions to NVSS 2146+82 (see \S 3.5
below) were obtained with the HYDRA multi-fiber positioner and the
Bench Spectrograph as part of the WIYN\footnote[6]{The WIYN Observatory
is a joint facility of the University of Wisconsin-Madison, Indiana
University, Yale University, and the National Optical Astronomy
Observatories.} Queue Experiment over the period of 14-22 September
1998.  The T2KC CCD was used as the spectrograph detector in its
spatially binned 1024 $\times$ 2048 pixel mode.  All exposures were
made with the red fibers, the Simmons camera, and a 400 lines/mm
grating.  The spectral resolution in this configuration is $\sim$4.5
\AA.

We calculated an astrometric solution for the KPNO 4-m frame of the
NVSS 2146+82 field using positions for stars in the frame taken from
the USNO A1.0 catalog (\cite{monet96}).   Using this solution, we
derived positions with the accuracy required by the HYDRA positioner
for our target galaxies.  Due to fiber placement restrictions and the
density of our target galaxies on the sky, we were only able to place
46 fibers on targets. The remaining 50 fibers were randomly placed on
blank sky, and they were used during the reduction process for night
sky subtraction.

The nine 30 minute program exposures were reduced using the IRAF
DOHYDRA script.  The weather conditions during the last two nights were
poor, and the spectra from these nights were not usable.
Therefore the final spectra were obtained by co-adding only the data
from nights one and two, a total of two hours of integration.

\subsection{Redshifts and Line Luminosities}

In Figure \ref{fig:oldfig2}, we present a contour plot of the V band
surface brightness from the central $40\arcsec \times 40\arcsec$ region
of the KPNO 4-m image after smoothing with a 3 pixel by 3 pixel boxcar
kernel.  Although we find that the elliptical galaxy at the radio core
position ($\alpha = 21^{\rm h}45^{\rm m}30^{\rm s}$, $\delta =
+81\arcdeg54\arcmin55\arcsec$ J2000.0) has a narrow line AGN emission
spectrum with a redshift of $z=0.145$, we find that the object
just to the east, which was assumed to be potentially a second nucleus,
has a zero-redshift stellar spectrum, indicating it is a foreground
star.  Figure \ref{fig:oldfig3} shows two plots of the wavelength and
flux calibrated spectrum of the host galaxy of NVSS 2146+82.

An unusual feature of the spectrum (Figure \ref{fig:oldfig3}) of the
AGN is that all of the emission lines appear to be double peaked.  The
second panel in Figure \ref{fig:oldfig3} shows an expanded view of the
$[{\rm O III}]$ doublet clearly showing the double peaked profile of
the emission lines.  Each emission line was easily fit with a blend of
two gaussians, indicating that AGN line emission is coming from two
sources with a velocity separation of $\sim$450 km s$^{-1}$.

Since the AGN emission line spectrum gives two different velocities, we
have decided to take the velocity of the stellar component of the
galaxy as the systemic velocity of the galaxy.   The stellar absorption
line redshift, calculated by cross-correlating the host galaxy spectrum
with the spectrum of the star immediately to the east, is
0.1450$\pm$0.0002.

Table \ref{table:linedata} lists the properties of the observed
emission features in the spectrum of NVSS 2146+82.  The redshifts of
the AGN emission line components were calculated by identifying
features and taking the average redshift of all of the identified
features.  In this way, the two AGN emission line components have been
measured to be at velocities of 40070$\pm$50 km s$^{-1}$ and 40520$\pm$50
km s$^{-1}$, which corresponds to redshifts of 0.1440$\pm$0.0002 and
0.1456$\pm$0.0002 respectively.  This indicates that the gas which is
giving rise to the bluer component of the AGN emission line spectrum
is moving relative to the stars in the AGN host galaxy at $-280$ km s$^{-1}$
and the gas emitting the redder lines is moving at 170 km s$^{-1}$ with
respect to the stars.

Each emission feature identified in Table \ref{table:linedata} was fit
with a blend of two gaussian components (except for the two
weak lines $[$\ion{Ne}{3}$]$ $\lambda3967$ and $[$\ion{O}{3}$]$
$\lambda4363$, where a single gaussian was used) to determine the line
flux.  The fluxes listed in Table \ref{table:linedata} were measured
after the spectrum of NVSS 2146+82 was flux calibrated using the
average of four measurements of the calibrator Feige 34.  The flux of
the calibrator varied significantly among our four separate exposures,
and we therefore estimate our spectrophotometry is only accurate to
about 20\%.  In addition to calibration error, there is an additional
error in the profile fitting, and therefore the errors listed for the
fluxes include both calibration and measurement error.

We derived an extinction of $A_{V} = 0.9\pm0.9$ (the large error is due
mostly to the calibration error in the fluxes) using the standard
Balmer line ratios for Case B recombination (\cite{osterbrock}) and the
extinction law of \cite{cardelli}.  The Galactic extinction at
the position of NVSS 2146+82 is given as $A_{V} = 0.5$ on the
reddening maps of \cite{schlegel}.  This value is consistent
with our Balmer line derived value, but possibly indicates that there
may be some dust in the host galaxy itself. We decided to correct the
measured line fluxes for reddening using the mean value we derived of
$A_{V} = 0.9$.  The errors listed in Table \ref{table:linedata} for
the fluxes do not include the error in the extinction determination.

\subsection{Optical Properties of the Host Galaxy}

\cite{sandage72} found that the
optical luminosity function of radio galaxy hosts was similar to that
of first ranked cluster members, and he noted that their optical
morphology was similar to bright E galaxies.  Although it was
therefore generally believed that the hosts of all radio galaxies were
gE types, subsequent large surveys of radio galaxies showed a good deal of
evidence for peculiar morphologies (e.g., \cite{heckman86}).  We
find that the host galaxy of NVSS 2146+82 is likely typical, i.e. it
is a gE galaxy, but with evidence of some peculiar morphological
features.

The broadband colors of NVSS 2146+82 are typical of bright FR II host
galaxies.  The absolute magnitude we derive for the host is $M_{V} =
-22.9$ at $z=0.145$ if we adopt a K correction of 0.46 magnitudes in
the V passband (\cite{kinney96}).  This magnitude is consistent
with the host being a gE galaxy, and also is very similar to the mean V
magnitude for 50 low redshift FR IIs of -22.6 (\cite{zirbel96}).

Similar to other FR II host galaxies, we find the optical morphology
of the host elliptical of NVSS 2146+82 to be disturbed.  In Figure
\ref{fig:oldfig2}, the four distinct objects besides the host galaxy
and foreground star have been identified as having non-stellar
morphologies with the Faint Object Classification and Analysis System
(FOCAS, \cite{valdes82}).  If these four galaxies share the same redshift
as the gE host of NVSS 2146+82, they all lie 50--100 kpc away from its
nucleus, a distance that implies that they may be dynamically
interacting with it.  Figure \ref{fig:oldfig2} also shows what appears
to be a bridge of diffuse optical light that almost connects NVSS
2146+82 to the galaxy to the southwest.  This bridge may indicate that
this smaller galaxy has recently passed close enough to NVSS 2146+82
to interact with it gravitationally.  There is also a fifth object $5\arcsec\:$
to the southeast of the center of NVSS 2146+82, which could
be in the process of merging with the gE galaxy.  However, due to the
faintness of this object and its proximity to the nucleus of 2146+82,
we are unable to classify this object definitively as a galaxy with
the FOCAS software.  Although we cannot conclude based on this image
that NVSS 2146+82 is undergoing a merger, its outer
isophotes do show evidence that it has been disturbed.

Correlations between the radio power and optical emission line
luminosities in radio galaxies have been established in several studies
(e.g., \cite{raw91}; \cite{zirbel95}; \cite{tad98}).  These
radio/optical correlations are assumed to arise primarily due to the
fact that both the radio jet and the ionization source originate in the
central engine.  The radio core power at 5 GHz ($\log P [W/Hz] =
23.85$) and the H$\alpha+[$\ion{N}{2}$]$ luminosity ($\log L [W] =
35.2$) for NVSS 2146+82 lie well within the dispersion in the
correlation in these quantities found for low redshift FR IIs
(\cite{zirbel95}).  This apparently indicates that the physical
conditions that cause this radio/optical correlation to arise may be
similar in this GRG and in ``normal'' FR IIs.

The shape of the emission line profiles in NVSS 2146+82 are not unique; 
emission line profiles and narrow band imaging of Seyfert galaxies and
radio galaxies have shown evidence for interaction between the radio
synchrotron emitting plasma and the optically emitting ionized gas (see
e.g., \cite{dmw89}).  Although the majority of objects that show
kinematic evidence for interactions between the radio jets and ionized
gas clouds tend to have more compact radio structures, the double
peaked line profiles seen in NVSS 2146+82 appear similar to those seen
in radio galaxies with jet/cloud interactions.  A recent model 
(\cite{tda92}) for interactions between the nuclear radio emission and
NLR gas in Seyferts produces $[$\ion{O}{3}$]$ profiles for objects near
the plane of the sky that are very similar to the double peaked profiles
seen in NVSS 2146+82.  The model of Taylor et al.\ (1992) produces double
peaks in the line profiles of objects oriented close to the plane of the sky 
because the emission lines are postulated to arise from gas that is 
being accelerated as a bowshock expands into the ionized medium surrounding the nucleus.
They model the bowshock as a series of annuli, and each annulus contributes
most of its luminosity at the two extreme radial velocities found along the
line of sight.  Although the specifics of the model of Taylor et al.\ (1992),
such as the discrete plasmon emission from the radio nucleus, may not necessarily
apply in the case of NVSS 2146+82, it suggests that the narrow line profiles
observed for this FR II (which is assumed to be very near the plane of the sky)
can be produced plausibly in a model where the ionized
gas is in a cylindrical geometry around the radio jet.  

Double peaked {\em broad} lines have been observed in radio galaxies (e.g.,
Pictor A [\cite{he94}]), however the model that is typically invoked to explain
the broad line profiles requires the radio galaxy to be oriented close to the
line of sight.  Since NVSS 2146+82 does not show a broad line component
and is unlikely to be oriented close to the line of sight, the accretion disk
model relied on to fit double peaked broad lines in AGN is probably unrelated
to the emission line profiles observed in NVSS 2146+82.

Although a jet/cloud interaction appears to be the most reasonable
explanation for the double peaked narrow emission lines observed in the
spectrum of NVSS 2146+82, it is also plausible that a gravitational
interaction between the FR II host galaxy and its nearest companions
may be the source of the $\sim$450 km/sec separation between the blue
and red emission line peaks.  Higher spatial resolution long slit
spectroscopy is necessary to determine which cause is more likely.

\subsection{Environment}

Deep CCD imaging of the region surrounding the host galaxy of NVSS
2146+82 has revealed a large number of nearby galaxies.  These galaxies
are near the limiting magnitude of the POSS/DSS images, so NVSS 2146+82
appears to lie in a sparsely populated region of the sky in the DSS.
However, photometry from the deeper Palomar 1.52-m images gives $-22
\lesssim M_{V} \lesssim -19.5$ for these nearby galaxies if they also
lie at $z = 0.145$, indicating a possible association with NVSS
2146+82.  In Figure \ref{fig:oldfig4}, we present a region of the KPNO
4-m image of NVSS 2146+82 that is 0.5 Mpc on a side and that has all
identified galaxies with $m_{v} \lesssim 21.3$ (corresponding to $M_{V}
\lesssim -19$ at $z=0.145$) circled.  These images do not go deep
enough to allow accurate identification and photometry of all galaxies
to $M_{V} = -19$, so this sample is not complete.  However, even though the
sample shown in Figure \ref{fig:oldfig4} is probably only complete to
$M_{V} \sim -20.5$, we have identified 34 candidate galaxies
surrounding NVSS 2146+82.

Although there are no previous identifications of the cluster around
NVSS 2146+82 (at $b=21\fdg5$, it is too close to the Galactic Plane to
have been included in the Abell [1958] catalog), there is a Zwicky
cluster to the north, with NVSS 2146+82 lying only $\sim\!5\arcmin$
south of the southern border of the Zwicky cluster.  The Zwicky cluster
2147.0+8155 (B1950.0 coordinates) is a compact group with 56 members
classified as ``extremely distant'' or $z > 0.22$ (\cite{zwicky}).
While this gives a redshift for the Zwicky cluster larger than that of
NVSS 2146+82, it is close enough to $z = 0.145$ ($< 400$ Mpc more
distant) that we may be seeing NVSS 2146+82 in projection against a
background rich cluster.

In September of 1998 WIYN/HYDRA spectra were obtained of 46 candidate
galactic companions of NVSS 2146+82 to determine their redshifts.  The
sample of 46 was selected in the following way: (1) We selected all
objects morphologically classified as galaxies in the KPNO 4-m frame by
FOCAS with aperture magnitudes $<21$, resulting in an initial sample of
205 galaxies.  (2) We divided this group into two subdivisions: the
first being all galaxies within 0.5 Mpc of 2146+82 in projected radius,
and the second being all those outside of the 0.5 Mpc radius.  
However, due to exposure time limitations, the available sample taken
from the 34 galaxies identified in Figure \ref{fig:oldfig4} within
0.5 Mpc of the host was reduced to the 17 brightest galaxies.  Fiber
placement restrictions allowed us to observe only 11 of these 17
galaxies. Objects
from the sample outside of the 0.5 Mpc radius from NVSS 2146+82 were
assigned to 35 of the remaining fibers, leaving about 45 fibers on blank
sky to allow accurate sky subtraction.  Unfortunately, as mentioned in
\S 3.2 above, the weather conditions during some of the queue observing
were poor, and this limited the success of the program.  There was enough
signal-to-noise to identify features in the spectra of only 24 of the
46 objects successfully.  We found that 7 of the 24 objects with good
spectra were actually misidentified stars.

Nonetheless, from the remaining 17 spectra of galaxies in the field
surrounding NVSS 2146+82, we were successful in identifying what we
believe to be a true cluster that contains the radio source host
galaxy.  Figure \ref{fig:oldfig5} presents an image with the 17
galaxies with measured redshifts marked.  The
positions, redshifts, and magnitudes for these objects are listed in
Table \ref{table:redshifts}.  A quality factor is assigned for each
redshift using the 0 (unreliable) to 6 (highly reliable) scale of 
\cite{munn97}. The quality is determined using: $q =
{\rm min}[6,{\rm min}(1,N_{def}),+2N_{def}+N_{prob}]$, where $N_{def}$ is the
number of spectral features that are accurately identified (less than
5\% chance of being incorrect) and $N_{prob}$ is the number of
spectral features that are probably correct (about a 50\% chance of
being correct).  If $q > 3$ is adopted as the requirement for a reliable
redshift, 5 of the 17 galaxies have unreliable redshifts.  The
histogram plotted in Figure \ref{fig:oldfig6} is a redshift
distribution for the 17 galaxies, and it shows that 50\% (6) of the
reliable redshifts fall in the range of $z = 0.135 - 0.148$, with 5 of
those having redshifts of $z = 0.144 - 0.148$.

Extrapolating the redshift distribution for the sample of galaxies
identified around NVSS 2146+82 from the redshift distribution of the 17
reliable galaxy spectra suggests that the 2146+82 cluster may be Abell
richness class 0 or 1.  Of course, the statistics are very uncertain.
Of the 11 galaxies within a projected distance of 0.5 Mpc of NVSS
2146+82 that were in the WIYN/HYDRA sample, redshifts were measured for
three of them.  Two of these have $z=0.144-0.145$, while the third has
$z=0.135$.  We identified features in 21 of the remaining 35 spectra
that were measured for objects outside of the projected 0.5 Mpc
radius.  We found that 7 were misclassified stars, and 3 of the 14
galaxies with reliable redshifts had $0.144 < z < 0.148$.  Abell's
(1958) richness criterion was based on the number of cluster galaxies
within the range $m_{3}$ to $m_{3} + 2$ ($m_{3}$ is the magnitude of
the third brightest cluster member). For the NVSS 2146+82 cluster,
$m_{3}$ should be $< 18.3$, since the third brightest galaxy of the 7
(which includes NVSS 2146+82) we have found at $z = 0.145$ has $m =
18.3$.  Of the 205 galaxies originally found in the KPNO 4-m field
containing NVSS 2146+82, 123 of these fall within the $m_{3}$ to $m_{3}
+ 2$ range used for estimating the Abell richness.  If we apply the
percentages above to this sample of 123 galaxies, then $37 \pm 13$
might be at the same redshift as NVSS 2146+82.  To this point, we have
been considering the cluster richness inside of 0.5 Mpc, for comparison
with the $N_{0.5}^{-19}$ richnesses of Allington-Smith et al.\ (1993)
and Zirbel (1997), and also within an area $\sim$3.8 Mpc on a side,
which is the size of the KPNO 4-m field at $z=0.145$.  However, we must
note that the original richness criterion for Abell class 1 clusters
was that 50 or more galaxies were contained in a radius of 3 Mpc for
$H_{0} = 50$ km sec$^{-1}$ Mpc$^{-1}$ (\cite{abell}).  A circle of
radius 3$h_{50}^{-1}$ Mpc at $z=0.145$ subtends 507 square arcminutes
on the sky, nearly twice the amount of area covered in our image.  If
the calculated optical richness from the 4-m image galaxy sample is
taken as a lower limit to the number of galaxies within an Abell
radius, the richness class of the group surrounding NVSS 2146+82
appears to be at least Abell class 0.

\section{X--ray Observations and Constraints} 

Richness class 0 clusters of galaxies typically have luminosities
with L$_{x} \approx 10^{43-45}$ ergs s$^{-1}$ (\cite{Ebeling98}),
while X-ray AGN range from L$_{x} \approx 10^{40-44}$ ergs s$^{-1}$
(\cite{Green92}), so a cluster or bright AGN will easily be seen with a
medium length exposure with {\it ROSAT}. NVSS 2146+82 was observed
with the {\it ROSAT} High Resolution Imager (HRI) between 1998
February 24 and 1998 March 13 for a duration of 30.3 ksec to search
for any hot gas that might be associated with the apparent overdensity of
galaxies or for an X-ray luminous AGN. 

The data were analyzed with the IRAF Post-Reduction Off-line Software
(PROS). The HRI data were filtered for periods of high background and
corrected for non-X-ray background, vignetting, and exposure using the
computer programs developed by Snowden (\cite{Plucinsky93};
\cite{Snowden98}).  After filtering, the live exposure was 29.8 ksec.
The resulting X-ray image was convolved with a gaussian beam with
$\sigma = 2\arcsec$ to recover diffuse X-ray emission. The contours of
the image are shown superposed on the DSS image in Figure
\ref{fig:xray}.

A few sources were visible near the edge of the field, but there seem
to be no significant sources of X-ray emission associated with any
optical or radio sources within the $20\arcmin$ extent of NVSS 2146+82 (Figure
\ref{fig:xray}). We derived upper limits on both the AGN or cluster
emission by extracting the X-ray counts from the corrected X-ray image
using circular regions centered on the host galaxy of $20\arcsec$ and
$2\farcm25$, respectively. The region sizes were chosen simply because
$20\arcsec$ represents the size of a typical HRI point source and
$2\farcm25$ is roughly 1-2 times the typical size of a cluster
core at the distance of the radio galaxy. The X-ray background was
determined by extracting the X-ray counts from an annulus of
$2.25-5\arcmin$ centered on the nucleus of the radio host and removing
3 point sources using $20\arcsec$ circular regions. We used PIMMS
(\cite{Mukai93}) to convert the HRI count rate into an unabsorbed flux
in the 0.1-2.0 keV band, assuming an emission model and a Galactic
photoelectric absorption column of $1.058\times10^{21}$ cm$^{2}$
(\cite{stark}). For the AGN, we assumed a power law with a photon
index, $\Gamma$, of 2.0 and derived an upper limit at the $90\%$
confidence level of $3.52\times10^{-14}$ ergs cm$^{-2}$ s$^{-1}$, or
$3.63\times10^{42}\ h_{50}^{-2}$ ergs s$^{-1}$ at the distance of the radio
galaxy. Similarly for the cluster, we assumed a Raymond-Smith thermal
emission spectrum characterized by $kT=1$ keV which yielded an upper
limit of $1.33\times10^{-13}$ ergs cm$^{-2}$ s$^{-1}$, or 
$1.37\times10^{43}\ h_{50}^{-2}$ ergs s$^{-1}$.

Unfortunately, our limit on the X-ray emission from the radio galaxy is
not very stringent. \cite{fabbiano84} studied the X-ray properties of
several 3CR radio galaxies with the {\it Einstein} Observatory. They
found that the FR II's radio and X-ray luminosities are strongly
correlated. Thus with a radio flux of 6.8 mJy at 5 GHz, NVSS 2146+82
should have a nuclear X-ray flux of a few times $10^{42}$ ergs s$^{-1}$. This
flux is comparable to our upper limit. Taking into account the
intrinsic scatter in the radio/X-ray correlation, our non-detection of
the AGN is quite reasonable.

Our upper limit on the X-ray emission from hot cluster gas provides a
much stronger constraint. Most Abell richness class 0 clusters have
X-ray luminosities of $\approx10^{43-45}$ erg/s (\cite{Ebeling98}).
Therefore any cluster of galaxies associated with the radio galaxy
must be either intrinsically weak in X-rays or must be 
poorer than our optical estimate.   \cite{wan96} studied the X-ray emission
of low-redshift FR II galaxies and found that poor clusters that contain FR II
sources are underluminous in X-rays compared to similar clusters that do not
contain FR IIs.  The median X-ray luminosity for low-$z$ clusters with FR IIs
was found to be $1.3\times10^{42} h_{50}^{-2}$ ergs s$^{-1}$ while it is 
$1.33\times10^{43} h_{50}^{-2}$ ergs s$^{-1}$ for a sample of low-$z$ clusters
without FR IIs (Wan \& Daly 1996).   Assuming that the group surrounding
NVSS 2146+82 is similar to that of other clusters found around low-$z$ FR IIs
and is underluminous in X-rays, the optical richness estimate is probably
correct.

\section{Discussion}

\subsection{Physical Properties of the Radio Source}

\subsubsection{Size and Luminosity}

Our observations of NVSS 2146+82 clearly show that it is an unusually
large FR II radio galaxy.  Its angular distance from the north lobe to
the south lobe gives an unusually large extent of $\theta=19\farcm5$.
For our assumed cosmology and our measured redshift of $z=0.145$, the
linear extent of the radio structure is 4$h_{50}^{-1}$ Mpc, placing it
in the Giant Radio Galaxy (GRG) class, which we define as sources
larger than 2$h_{50}^{-1}$ Mpc.  NVSS 2146+82 is therefore the second
largest FR II known, surpassed only by 3C236 which is
$\sim$6$h_{50}^{-1}$ Mpc in extent.  FR II galaxies of this size are
extremely rare; a literature search by \cite{nilsson} of 540 FR IIs
contains only 27 objects with sizes greater than 1$h_{50}^{-1}$ Mpc.
Of this sample of 27 large FR IIs, only 5 are larger than
2$h_{50}^{-1}$ Mpc.  For comparison, the other known giant radio
sources are shown in Table \ref{table:giants}. The log radio luminosity
of NVSS 2146+82 at 1.4 GHz is 25.69, in the middle of the range for
giant radio sources.

It remains unclear if there are fundamental differences between GRGs
and ``normal'' radio galaxies.  The relative paucity of known GRGs may
be in part due to observational selection effects in past radio
surveys.  An alternative reason for the rarity of giant radio galaxies
may be that the physical conditions necessary for the creation of a GRG
are uncommon in the universe.  Although the similarity between NVSS
2146+82 and other FR IIs suggests that it is a typical FR II radio galaxy
at the extreme end of the size distribution, a study of a complete
sample of radio galaxies that includes GRGs will have to be made to
determine if GRGs are part of a continuous distribution in size of
normal radio galaxies or if there are fundamental differences between
GRGs and smaller FR IIs.

\subsubsection {Equipartition calculations}

If the usual equipartition assumptions are made, then it is possible to
estimate the magnetic field strength and pressure in the lobes.
Assuming that the observed spectral index is maintained from 10 MHz to
100 GHz, that there are equal energies in the radiating electrons and
other particles, and that the filling factor is unity, the derived
magnetic field is B$_{min}\ \approx\ 5\times 10^{-6}\ h_{50}^{2/7}$ Gauss
and p$_{min}\ \approx\ 3.5\times 10^{3}\ h_{50}^{4/7}$ cm$^{-3}$K for the
hot spots. At the midpoint of the lobes these values are
B$_{min}\ \approx\ 8\times 10^{-7}\ h_{50}^{2/7}$ Gauss and
p$_{min}\ \approx\ 2.3 \times 10^{2}\ h_{50}^{4/7}$ cm$^{-3}$K. At this
redshift, the 3~K microwave background has an equivalent magnetic field
of 4.2$\times 10^{-6}$ Gauss so the energy loss in the lobes should be
dominated by inverse Compton scattering of this background, and the
time for the electrons radiating at 1400 MHz to lose half of their
energy will be $\approx\ 10^{8}\ h_{50}^{-3/7}$ years.

\subsubsection {Magnetic Field and Faraday Rotation}
\begin{figure}
\plottwo{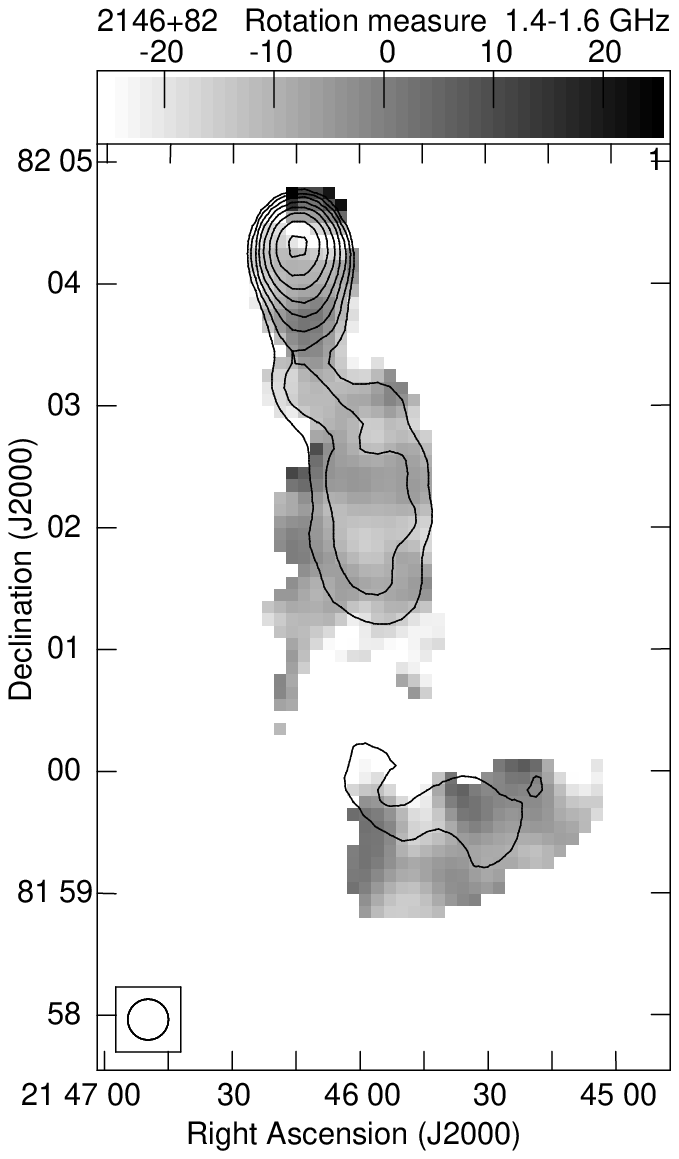}{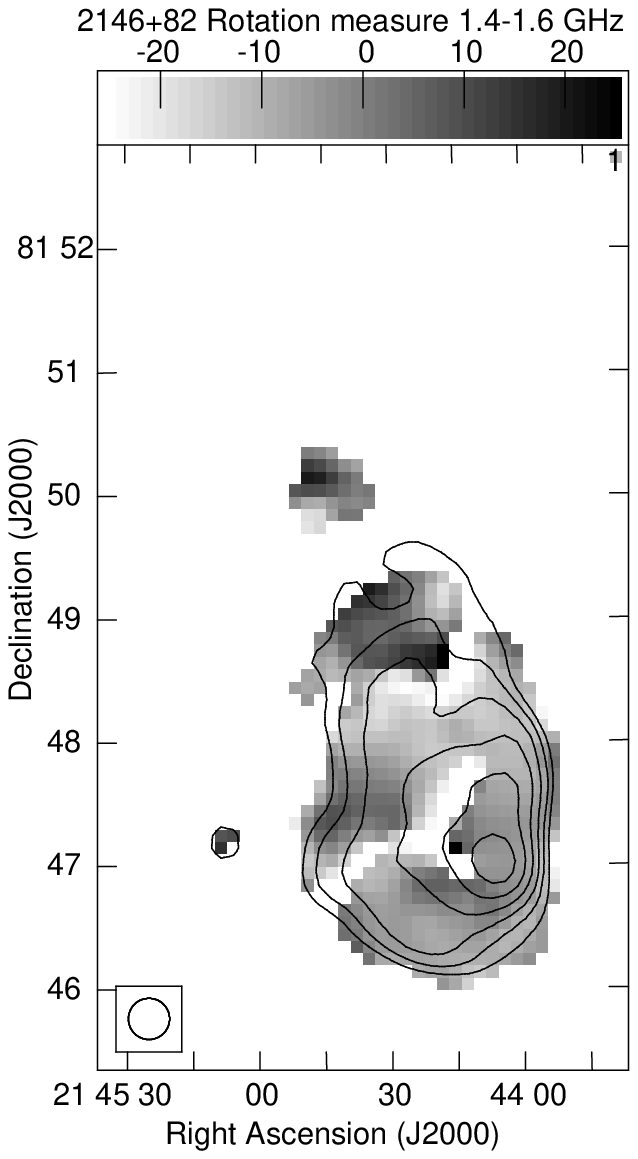}
   \caption{Gray scale representation of the rotation measure at 20\arcsec\
    resolution with superimposed contours of the 1.6 GHz total
    intensity at the same resolution.
    The bar at the top gives the grayscale values and the resolution
    is shown in the lower--left corner.
    The north lobe is shown in the left and the south lobe on the right.
    } 
    \label{fig:RM}
 \end{figure}

\begin{figure}
\plotfiddle{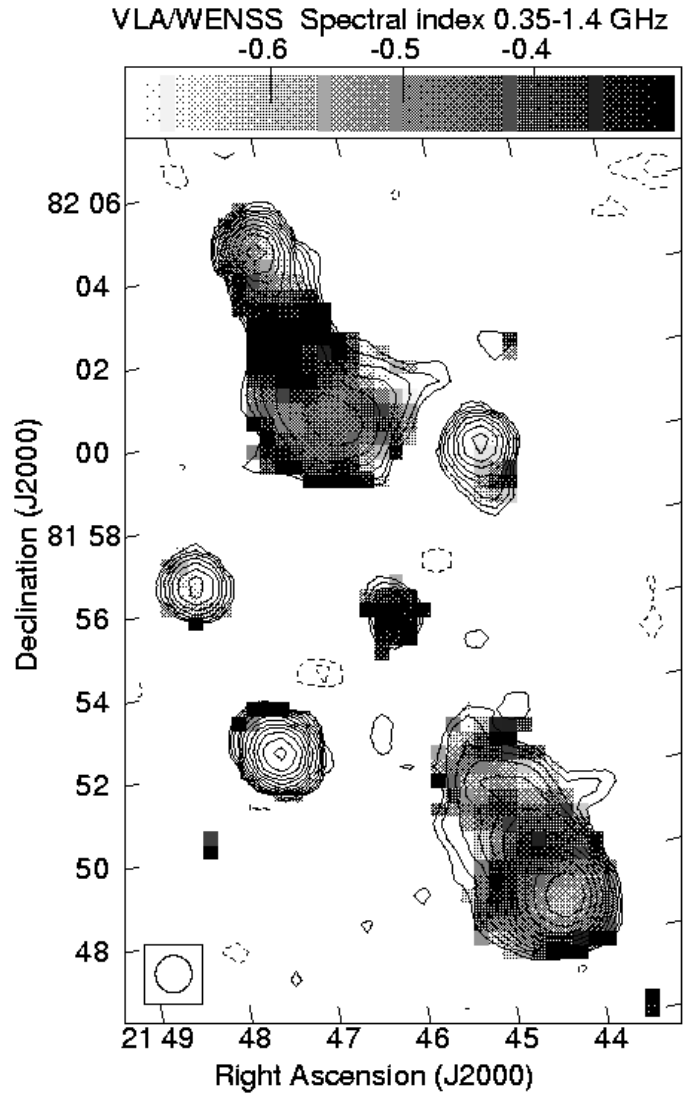}{6truein}{0}{70}{70}{-220}{-50}
 	\caption{Gray scale representation of the spectral index
        distribution derived from VLA measurements at 1.4 GHz and the
        0.35 GHz WENSS image with superimposed contours from the VLA image.
	The resolution is 54\arcsec\ (FWHM), illustrated in the
        lower--left and the bar at the top gives the gray scale values.
 	}
   \label{fig:WENSS}
 \end{figure}

\begin{figure}
\plottwo{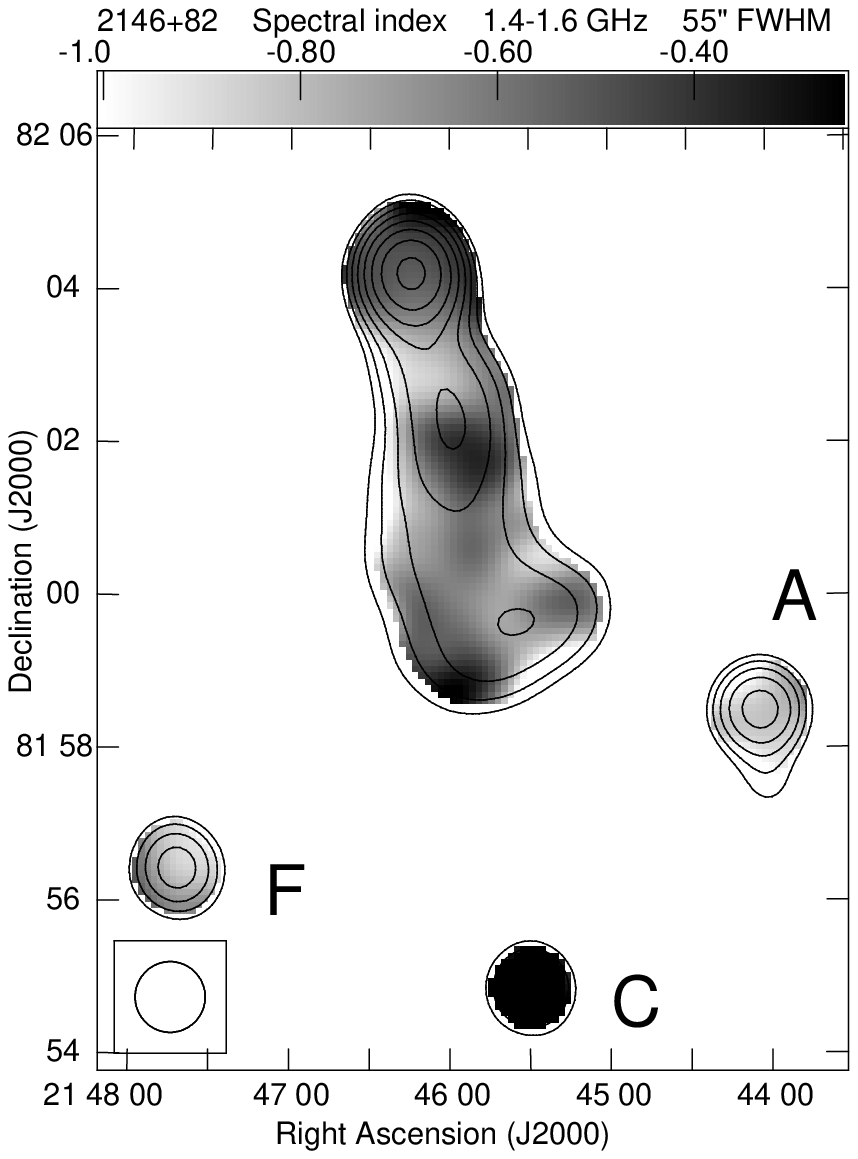}{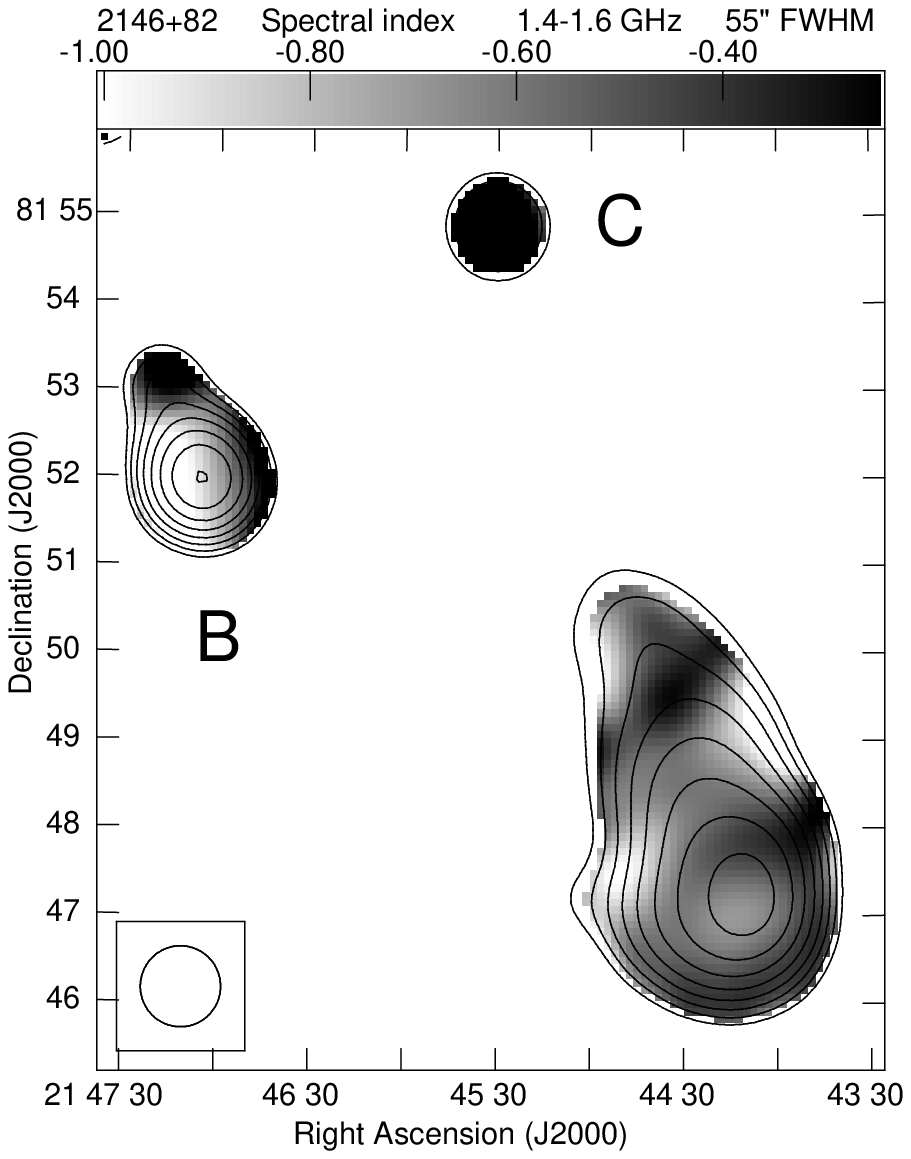}
   \caption{Gray scale representation of the spectral index at
    55$\arcsec\ $ resolution derived from the 1.36 and 1.63 GHz data
    with superimposed contours of the 1.63 GHz total intensity at the
    same resolution. 
    The bar at the top gives the gray scale values and the resolution
    is shown in the lower--left corner.
    The north lobe is shown in the left and the south lobe on the right.
    } 
    \label{fig:SI}
 \end{figure}

\begin{figure}
\plotone{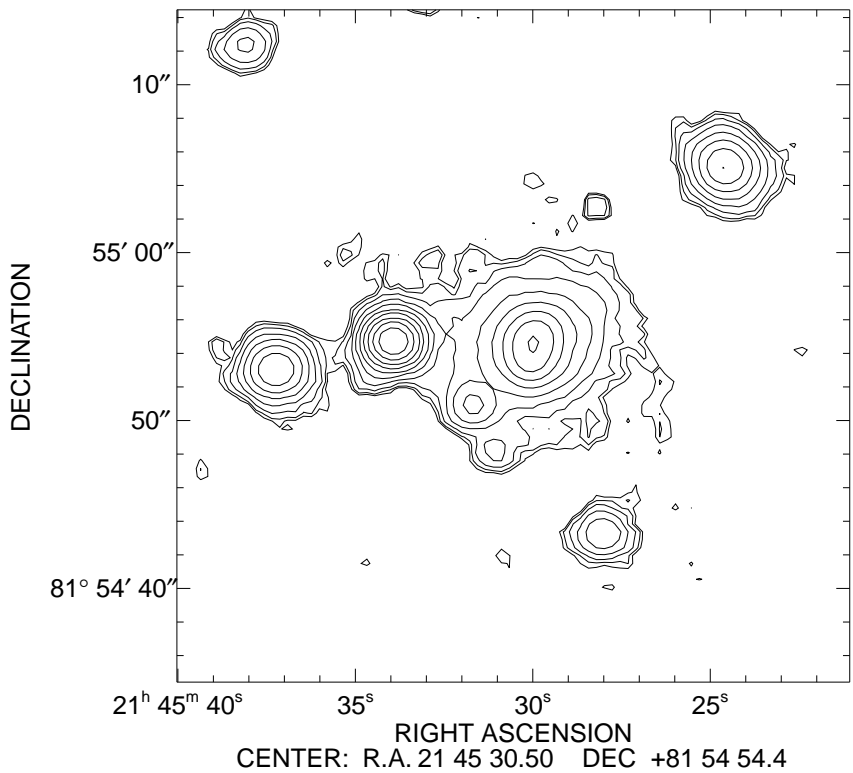}
\caption{A contour plot of the V band surface brightness of
the region immediately surrounding the host galaxy of NVSS 2146+82.
The object just east of the host galaxy (at center) is a foreground
star.  The remaining four discrete objects all have non-stellar PSFs,
indicating that they are very likely galaxies.  The object to the
northwest of NVSS 2146+82 is a galaxy and has a spectroscopic redshift
from our WIYN program of $z = 0.144$. }
\label{fig:oldfig2} 
\end{figure}

\begin{figure}
\plottwo{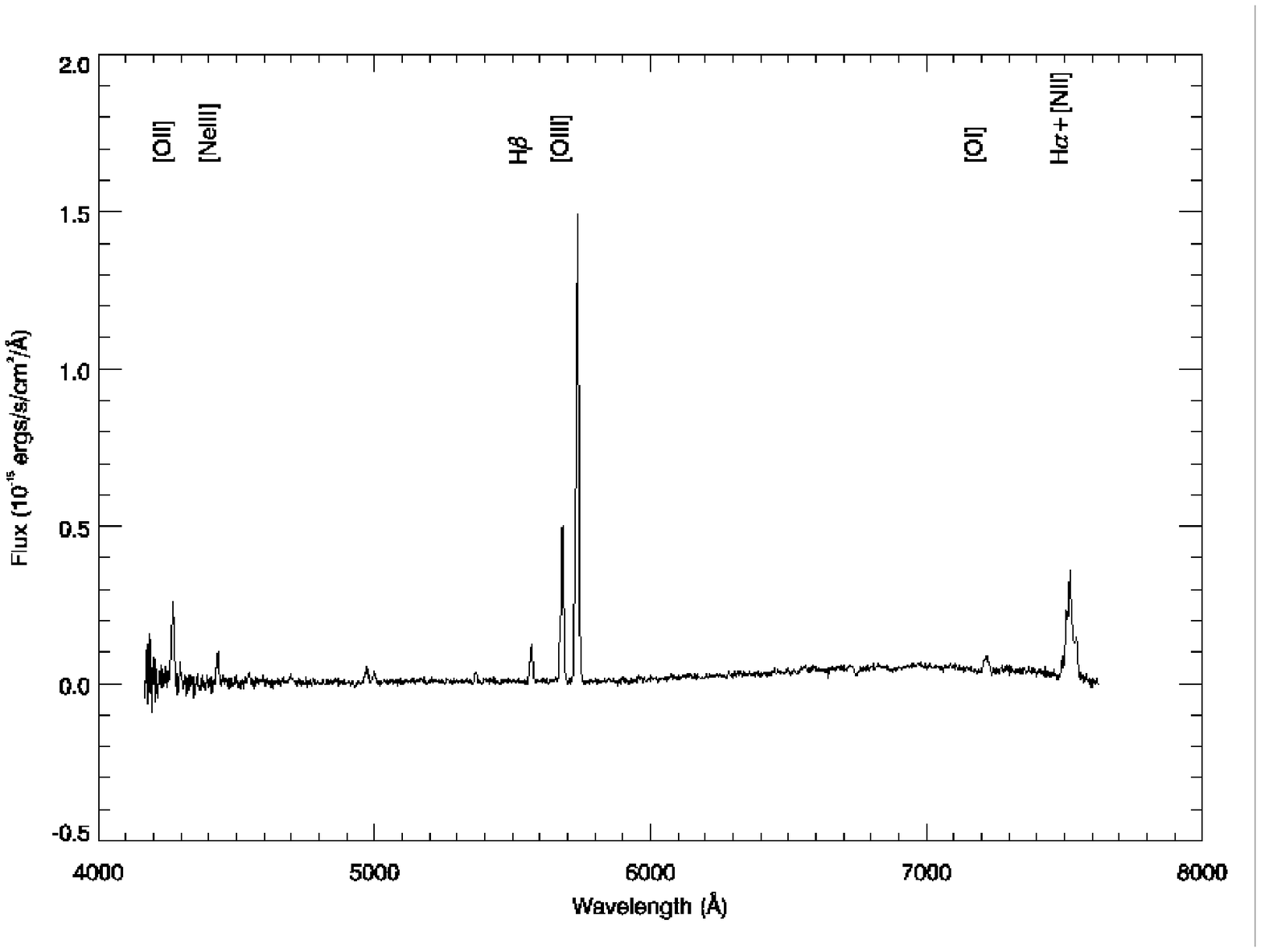}{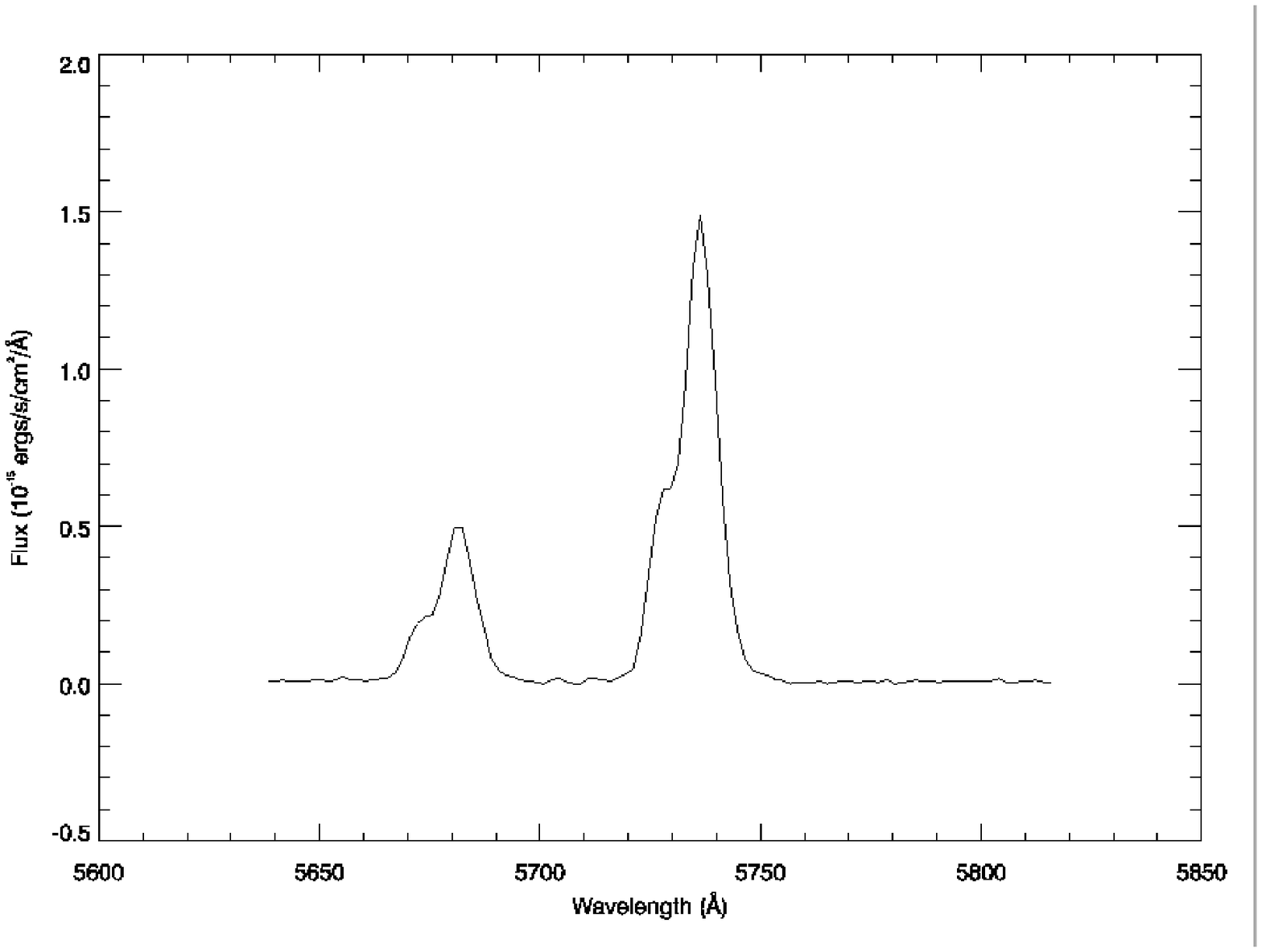}
\caption{Spectrum of the host galaxy of NVSS 2146+82.  The
left panel shows the full spectrum, with several of the stronger
emission features identified in Table \ref{table:linedata} are marked.
Most of the 
emission lines have a double-peaked profile, as illustrated in the
right panel with the $[$\ion{O}{3}$]$ $\lambda\lambda 4959$, 5007 pair.}
\label{fig:oldfig3} 
\end{figure}

\begin{figure}
\plotone{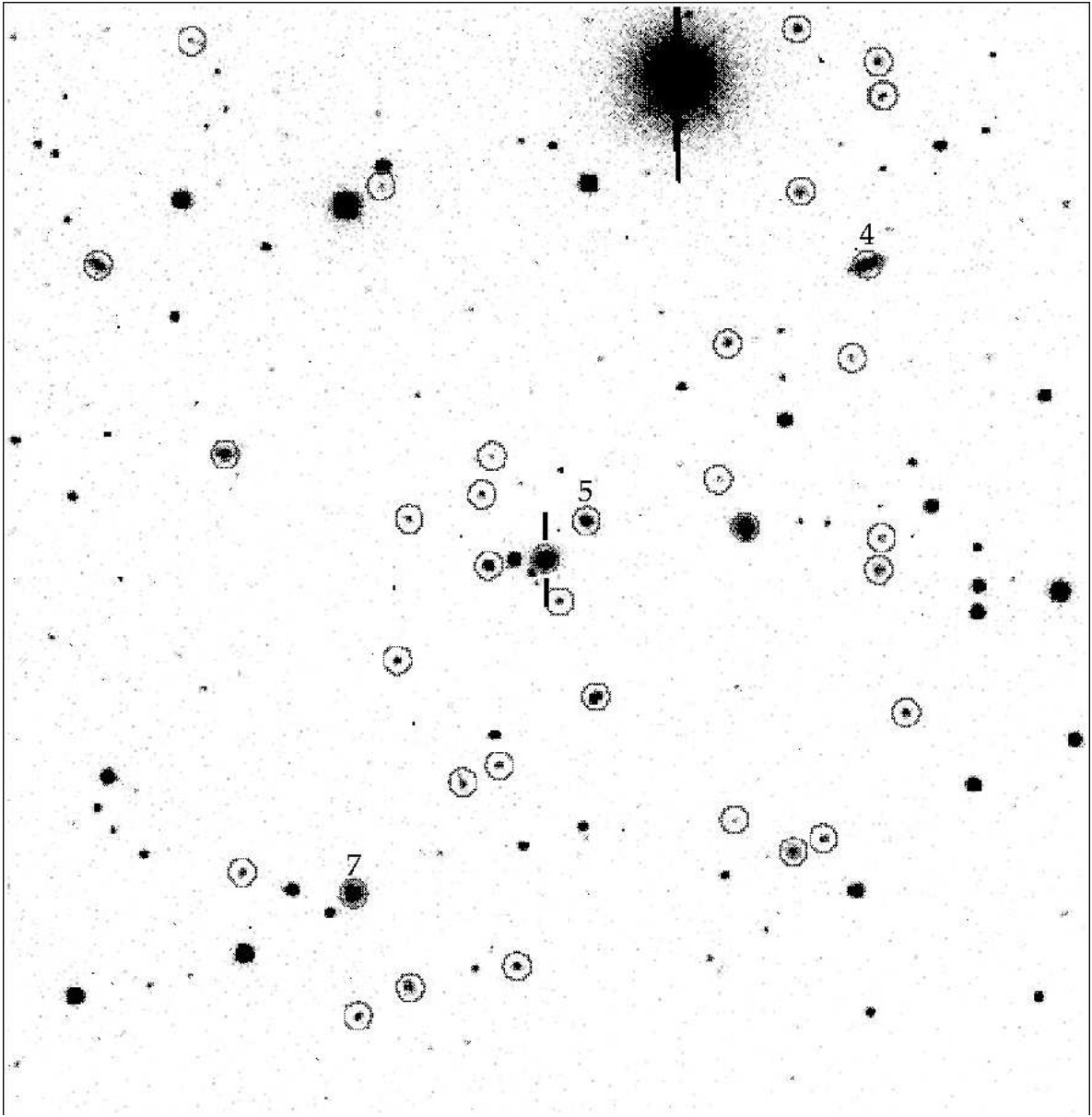}
\caption{A view of the field surrounding NVSS 2146+82 from
the central region of our KPNO 4 meter image (north is up, east to the
left).  This field is 0.5 Mpc on a side at the redshift of NVSS
2146+82, and it contains 34 objects down to $m_{v} = 21.3$ ($M_{V} \geq
-19$ at $z = 0.145$) that are morphologically identified as galaxies.
For reference, the host galaxy of NVSS 2146+82 is marked with hash
marks, and the three galaxies in this region that we measured
spectroscopic redshifts for are marked with their ID numbers from
Table \ref{table:redshifts}}
\label{fig:oldfig4} 
\end{figure}

\begin{figure}
\plotone{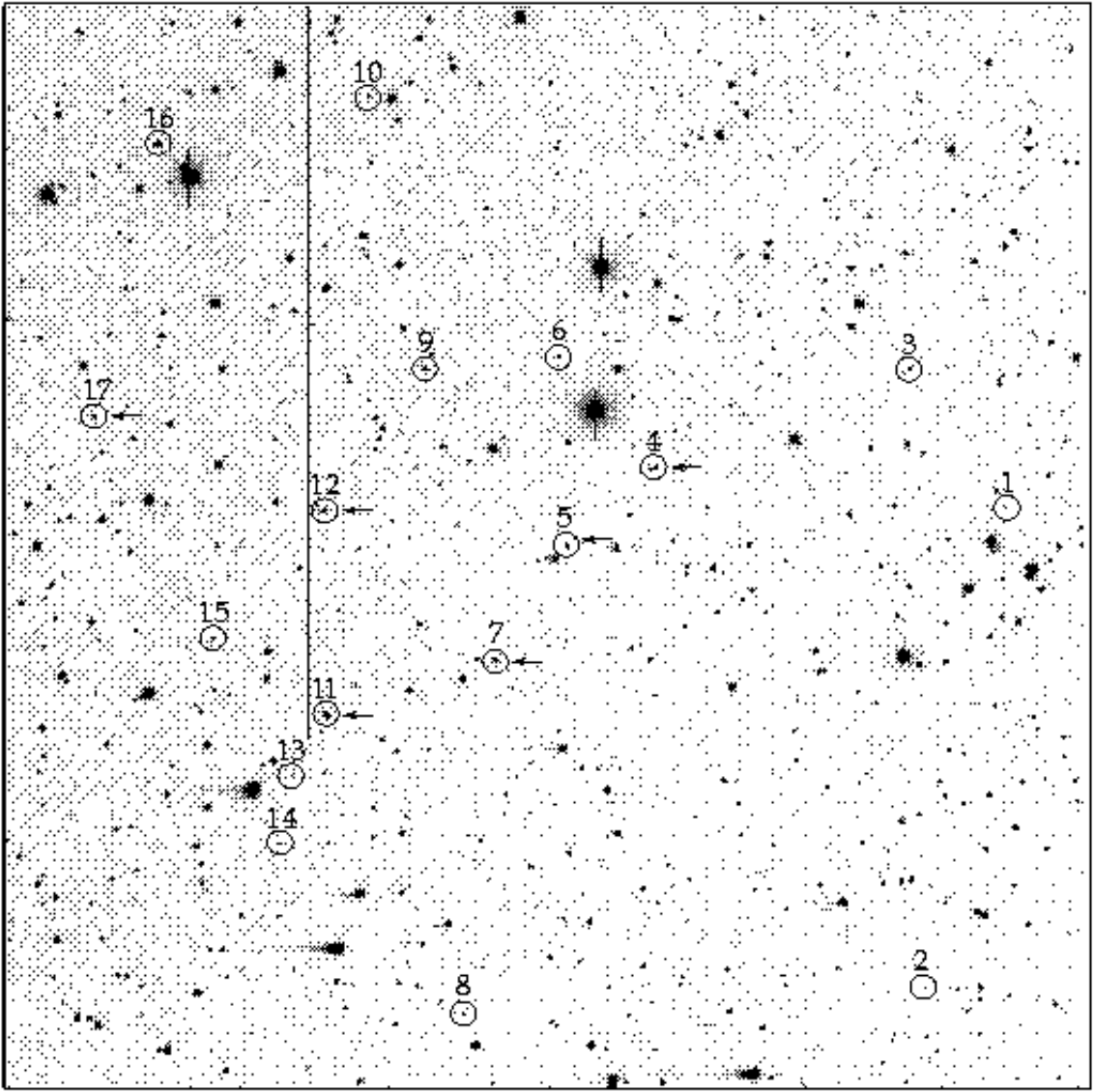}
\caption{The full field that we observed with the KPNO 4
meter surrounding NVSS 2146+82.  In this image, the 17 galaxies with
spectroscopic redshifts are circled and identified with their ID number
from Table \ref{table:redshifts}.  Those objects with reliable
redshifts in the range 
$0.135 < z < 0.149$ are marked with arrows.  NVSS 2146+82 is the galaxy
just outside of the southeast edge of the circle surrounding galaxy 5.}
\label{fig:oldfig5} 
\end{figure}

\begin{figure}
\plotone{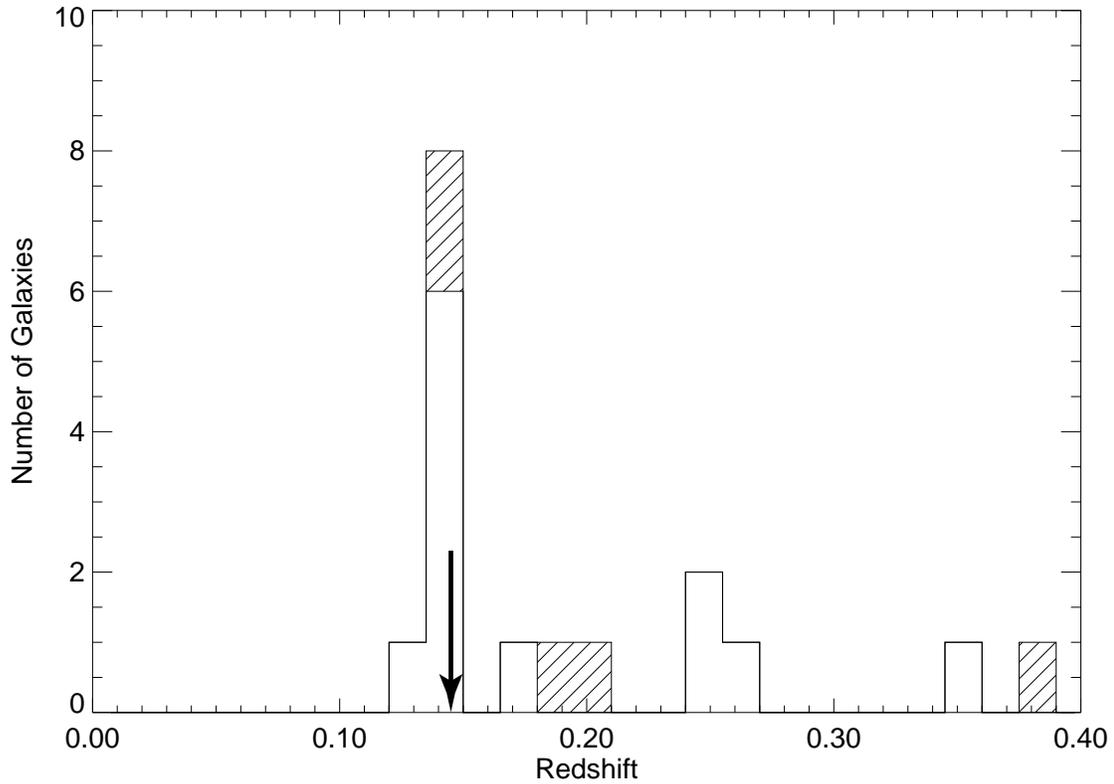}
\caption{A histogram of the redshifts of the 17 galaxies
that we obtained spectra for with the WIYN.  The empty histogram is the
distribution of redshifts that have $q > 3$, and the hatched histogram
is the distribution of the lower quality redshifts.  The arrow shows
the redshift for NVSS 2146+82, $z=0.145$.  The peak in this diagram is
centered around $z=0.1425$, showing that 6 -- 8 galaxies in our sample
of 17 share the same redshift as NVSS 2146+82.}
\label{fig:oldfig6} 
\end{figure}

\begin{figure}
\plotone{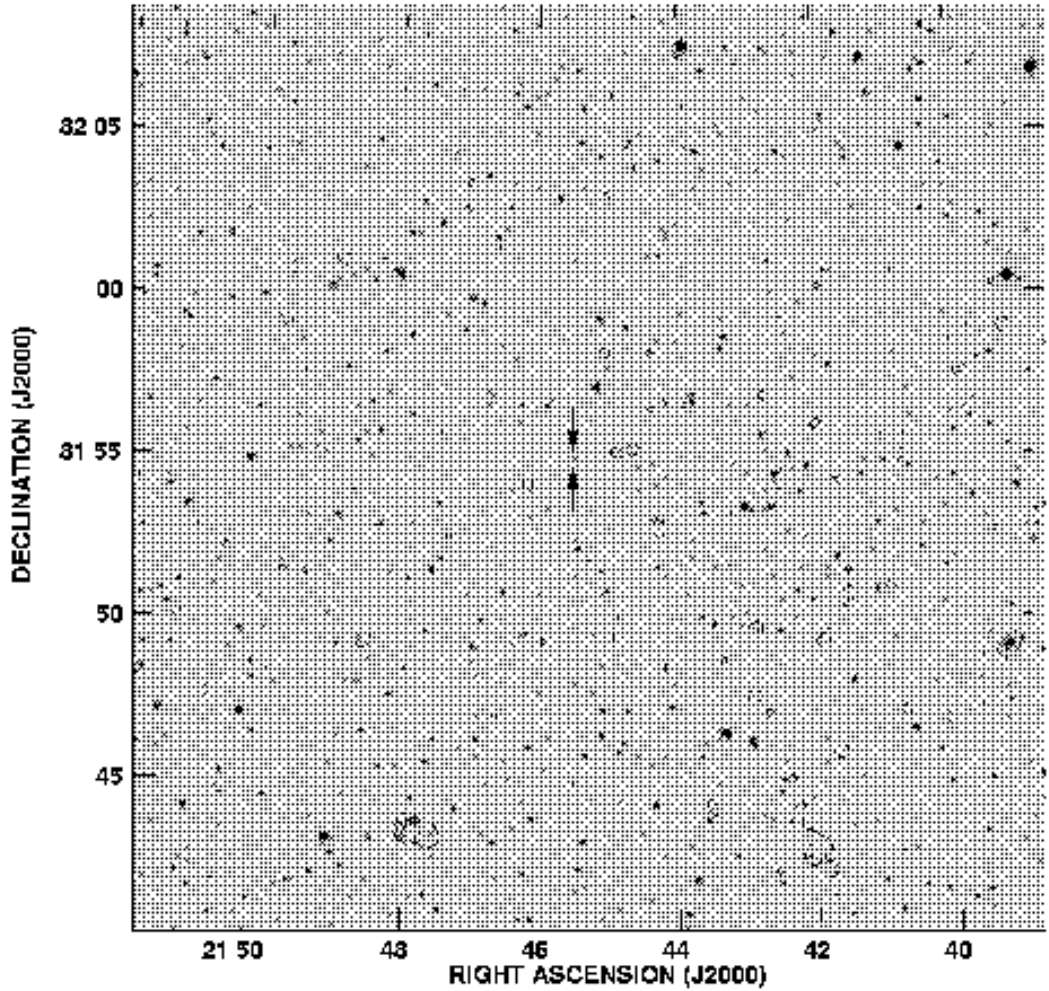}
\caption{A contour plot of the {\it ROSAT} HRI X-ray image
in the 0.5--2.0 keV band. The X-ray image has been corrected for
non-X-ray background, vignetting, and exposure and convolved with a 2
arcsec sigma gaussian beam. The contours are superposed on an optical image
from the Digitized Sky Survey (Lasker et al.\ 1990).  The base contour
level is 1.1 counts$/$pixel. The contours plotted are multiples ($1$,
$2^{1/2}$, $2^{1}$, $2^{3/2}$, ...) of the base contour level. The
arrows indicate the position of the host galaxy.}
\label{fig:xray}
\end{figure}

The mean Faraday rotation of $\approx$ $-9$ rad\ m$^{-2}$ shown in
Figure \ref{fig:RM} is consistent with the results of
\cite{Simon-Normandin} for other extragalactic sources seen through 
this region of the Galaxy ($l=116.^\circ7, b=21.^\circ5$). It is
therefore likely that the rotation measure screen seen in Figure 
\ref{fig:RM} is primarily the foreground screen of our Galaxy. 
The low apparent rotation measure and the smoothness of the
polarization structure shown in Figure \ref{fig:polarization} suggests
that the magnetic field in this source is well ordered.  The field
configuration is entirely typical of older extended FR II sources, with
the E vectors lying approximately perpendicular to the ridge line of
the radio emission in most features.

We note that the greater variance and evidence for organized
structure in the Faraday rotation of the southern lobe is the
opposite of what would be expected if the jet sidedness were due to
Doppler favoritism and the Faraday rotating medium were local to
the source.  We think it more likely that the Faraday rotation
structure arises along the line of sight in our Galaxy.

\subsubsection {Spectral Index Variations}
The spectral index variations shown in Figure \ref{fig:SI} indicate
that there are regions $2\farcm4$ back towards C from the brightest
region in each lobe that have unusually flat spectra
($\alpha^{1.4}_{0.35}\ \approx$ -0.3), flatter even than the hot spots.
The only extended synchrotron sources known with spectra this flat are
a few Galactic supernova remnants (\cite{Berkhuijsen}).

The spectral index structure in NVSS 2146+82 is unlike the systematic
steepening of the spectrum away from the hot spots that is usually
interpreted as an effect of spectral aging in extended lobes.  In such
interpretations, electrons are presumed to be injected into a high
field region in or around the hot spots, and their energy spectrum
steepens with distance as they diffuse into lower field regions of the
extended lobes.  Clearly no such interpretation can be made here.

These flatter spectrum regions occur in the transition zone from the
featureless parts of the lobes (closer to the core) to the parts near
the regions of enhanced emission that contain significant filamentary
structure.  The anomalous regions are near the midline of the lobes;
the southern region is centered on the path of the jet and the northern
region is at one end of a prominent filament (the path of the jet is
uncertain).  The relative symmetry of the flatter spectrum regions of
the lobes suggests that they might be produced by an intrinsic property
of the source, such as a variable spectral index in the injection
spectrum of the relativistic electrons from the jet, rather than local
environmental effects.

If the magnetic field has values near those
estimated by the equipartition calculations given above, then the
energy loss of the radiating electrons is dominated by inverse Compton
scattering against the Cosmic Microwave Background.  In the low
density, low magnetic fields in these lobes, the aging effects will be
slow and the history of a variable electron spectrum could be
maintained along the length of the lobe.

\subsubsection{Size Scales of Symmetry in the Radio Source}

There are three size scales on which symmetries appear or change in the
radio structure:  The first is $1\farcm5 = 300 h_{50}^{-1}$ kpc.  The
jets appear to become symmetric on this scale but are asymmetric on
smaller scales.   If the J2 and K components (Figure 4) are symmetric
features in the jet and counterjet, any Doppler boosting from
relativistic motion must have disappeared by this point in the jet.
The second scale is $3\farcm2 = 640 h_{50}^{-1}$ kpc.  On this scale,
there is a dramatic brightening of both lobes.  The third scale is
$6\farcm5 = 1300 h_{50}^{-1}$ kpc.  At this distance, the lobes become
even brighter and strong filamentary structure appears.  This is the
distance at which regions of spectral anomaly appear in the extended
emission.

The largest scale symmetries thus suggest a symmetric overall
environment, apart from the slight non-collinearity (C-symmetry) of the
structure.  The small scale brightness asymmetries of the jet and
counterjet might be attributed to Doppler boosting and dimming by
relativistic motion which effectively disappears by $\sim 300
h_{50}^{-1}$ kpc, i.e. on a scale more typical of a ``non-giant''  FR
II source.  We reiterate however that the small asymmetry in rotation
measure dispersion (variance) between the lobes is opposite in sign to
that expected on this interpretation. This asymmetry seems more likely
to reflect an intrinsic asymmetry (or gradient) in the foreground
magnetoionic medium.

\subsection{The Optical Environment}

One possibility for the origin of GRGs is that they are
otherwise normal FR II sources that reside in extremely low
density gaseous environments. 
The environments in which radio galaxies reside have been studied in
depth (e.g. \cite{longair79}; \cite{heckman86}; \cite{prestage88};
\cite{hill91}; \cite{aezo}; \cite{zirbel97})
because the gas density and pressure in the host galaxy's ISM,
any intracluster medium, and the IGM are at least partly responsible
for determining the resulting radio morphology. 

An intriguing result of recent studies (\cite{hill91};
\cite{aezo}, \cite{zirbel97}) is that FR II galaxies are
found in a range of cluster richnesses at moderate redshifts, but they
are only found in poor to very poor groups at low redshift.  The
``richness'' of the cluster associated with a radio galaxy can be
estimated in a statistical sense in the absence of redshift data on
nearby galaxies.  Allington-Smith et al. (1993) define the richness
parameter $N_{0.5}^{-19}$ as the number of galaxies within a projected
radius of 500 kpc and with $M_{V} \leq -19.0$ assuming the same redshift
as the AGN.  The number counts are corrected for contamination by
foreground and background galaxies by subtracting number counts from a
field offset from the radio galaxy.  Zirbel (1997) gives a conversion
of $N_{0.5}^{-19}$ to Abell class as $N_{Abell} = 2.7
(N_{0.5}^{-19})^{0.9}$.  With this conversion, the thresholds for
Abell Classes 0 and 1 are $N_{0.5}^{-19} = 15$ and 26 respectively.
Using this richness estimation technique, Zirbel (1997) found that of
a sample of 29 low redshift ($z < 0.2$) FR IIs: (1) 41\% of the sample
of low $z$ FR IIs reside in very poor groups ($N_{0.5}^{-19}< 3.5$),
and (2) more importantly, no low redshift FR II was found in a rich
group with $N_{0.5}^{-19} > 20$.  Based on the results given in \S 3.5,
NVSS 2146+82 appears to reside in a group with an anomalously high
galaxy richness compared to other low redshift FR IIs.  Although
the galaxy counts from the field surrounding NVSS 2146+82 were not
calculated identically to those of Zirbel (1997), the value of
$N_{0.5}^{-19}$ is likely $> 25 - 30$ for NVSS 2146+82.

The upper limit on the cluster X-ray emission is consistent with the
NVSS 2146+82 group being at the low end of the X-ray luminosity 
distribution for poor clusters.  \cite{wan96} found that 
in a comparison of low redshift clusters with and without FR II
sources, clusters that contained FR IIs were underluminous in X-rays
compared to clusters without FR IIs.   Although the cluster surrounding
NVSS 2146+82 may be Abell Class 0, its lack of associated X-ray gas
suggests that the pressure in the surrounding medium is low enough
for a giant radio source to form with little disruption of the FR II jet.

Curiously, several other GRGs listed in Table \ref{table:giants} also
appear to lie in regions with overdensities of nearby galaxies.  The
GRG 0503-286 appears to lie in a group of 30 or so galaxies
(\cite{sar86}).  These companions are concentrated to the northeast of
the host galaxy of 0503-286, and may have caused the asymmetric
appearance of the northern lobe of the radio structure.   Overdensities
of nearby galaxies are also reported for 1358+305 (\cite{par96}) and 8C
0821+695 (\cite{lacy93}); however, in both cases there is no
spectroscopic confirmation of the redshifts of the candidate cluster
galaxies.  In a recent study of the optical and X-ray environments of
radio galaxies, \cite{miller99} find that for a sample of FR I
sources, all have extended X-ray emission and overdensities of optical
galaxies.  However of their sample of seven FR II sources, none have
overdensities of optical galaxies or extended X-ray emission except for
the GRG DA240, which has no extended X-ray emission
but does have a marginally significant excess of optical companions.
Perhaps for at least some of the GRG population, the presence of the host
galaxy in an optically rich group with little associated X-ray gas is
related to the formation or evolution of the radio source?

\section{Summary and Conclusions}

We have presented multi-wavelength observations of the
unusually large FR II radio galaxy NVSS 2146+82.  The overall size of the radio
source is $4h_{50}^{-1}$ Mpc, making it the second largest known
FR~II source.  We have found the host galaxy to be similar in both
luminosity and morphology to a sample of other low redshift FR II
galaxies.  Emission line profiles seen in the spectrum of the host
galaxy are double peaked, which may indicate that the ionized gas may
be being accelerated by the bipolar radio jet.

We have also found evidence for an anomalously rich group of 
galaxies at the same redshift as NVSS 2146+82 that has little
associated X-ray emitting gas.  Though unusual in having a rich
environment, this source is similar to other low redshift FR IIs in clusters;
the NVSS 2146+82 group is underluminous in X-rays compared to clusters
of similar richness that contain no FR II.  The large radio size, lack
of significant Faraday rotation and non detection of X-rays all suggest
that in spite of the richness of the cluster in which this galaxy
resides, it has a low gas density.

There is some morphological evidence that the host galaxy of NVSS 2146+82
may be undergoing tidal interaction with one or more of its nearest
companions.  Also, an interaction may be responsible for the double-peaked
emission line profiles, however the spatial resolution of the spectrum
of the nucleus is not high enough to distinguish between a merger 
origin or radio jet/cloud interaction origin for the peculiar profiles.

Apart from the radio spectral index anomaly, the radio properties of 
this source are like a normal FR II
source scaled up by a factor of ten, preserving the standard overall
morphology and polarization structure.  In the outer regions of the
source the magnetic field is likely to be so weak that inverse
Compton losses to the Cosmic Microwave Background dominate synchrotron
losses.

\acknowledgements

We are grateful to Mark Whittle for many helpful conversations.  We are
grateful to Matt Bershady, Randy Phelps, and Mike Siegel for either sharing
observing time or taking observations in support of this research.
CP acknowledges the support of a Grant-in-aid of Research from Sigma
Xi, the Scientific Research Society.  This research has made use of the
NASA/IPAC Extragalactic Database (NED) which is operated by the Jet
Propulsion Laboratory, California Institute of Technology, under
contract with the National Aeronautics and Space Administration.  We
acknowledge the use of NASA's {\it SkyView} facility
(http://skyview.gsfc.nasa.gov) located at NASA Goddard Space Flight
Center.

\clearpage
\pagestyle{empty}

\clearpage

\pagestyle{empty}

\begin{table}
\caption[VLA Observing Log]{VLA Observing Log}
\begin{center}
\begin{tabular}{cccccc}

\tableline
\tableline
VLA           & Observing & Center Frequencies & Bandwidth & Number & Integration \\
Configuration &   Date   & (MHz)              & (MHz)     &  of Fields & (min)    \\
\tableline
A  &  1995 Jul 08 & 8415, 8465  &  50  &  6  &  5 \\
A  &  1995 Jul 08 & 4835, 4885  &  50  &  6  &  6 \\
B  &  1995 Dec 23 & 1365, 1636  &  12.5&  3  &  13 \\
C  &  1996 Feb 15 & 1365, 1636  &  25  &  3  &  22 \\
D  &  1996 Sep 02 & 1365, 1636  &  25  &  3  &  18 \\
B  &  1995 Dec 23 & 327.5, 333  &  3.1 &  1  &  69 \\
C  &  1996 Feb 15 & 327.5, 333  &  3.1 &  1  &  30 \\
D  &  1996 Sep 02 & 327.5, 333  &  3.1 &  1  &  7.5 \\
BnC & 1997 Jun 17 & 1365, 1435  &  50  &  3  &  185 \\
\tableline
\end{tabular}
\end{center}
\label{VLAObsLog}
\end{table}

\clearpage

\begin{table}
\label{Flux Densities}
\caption[Flux Densities]{Flux Densities}
\begin{center}
\begin{tabular}{cccccc}

\tableline
\tableline
  & 0.35 GHz\tablenotemark{a}  & 1.4 GHz            &  4.9 GHz & 8.4 GHz &  $\alpha^{1.4}_{0.35}$ \\
\tableline
Total   &  0.99 $\pm$ 0.02 Jy & 0.53 $\pm$ 0.1 Jy  &          &
&  -0.45 $\pm$ 0.06   \\
N lobe  &  0.43 $\pm$ 0.01 Jy & 0.24 $\pm$ 0.05 Jy &          &
&  -0.42 $\pm$ 0.07   \\
S lobe  &  0.52 $\pm$ 0.01 Jy & 0.27 $\pm$ 0.05 Jy &          &
&  -0.47 $\pm$ 0.06   \\
C       &  23 $\pm$ 2  mJy    & 13.6 $\pm$ 0.5 mJy & 6.8 $\pm$ 0.2 mJy
&  3.4 $\pm$ 0.2 mJy  &  -0.38 $\pm$ 0.03  \\
J1       &                     & 1.3  $\pm$ 0.2 mJy &          &        &      \\
J2       &                     & 0.3  $\pm$ 0.2 mJy &          &        &      \\
J3       &                     & 0.8  $\pm$ 0.2 mJy &          &        &      \\
K        &                     & 0.7  $\pm$ 0.2 mJy &          &        &         \\
\tableline
\end{tabular}
\end{center}
\tablenotetext{a}{ 0.35 GHz measurements are from the WENSS image.}
\label{table:fluxes}
\end{table}

\clearpage

\begin{table}
\caption[Mean Aperture Magnitudes for NVSS 2146+82 Host Galaxy]
{Mean Aperture Magnitudes for NVSS 2146+82 Host Galaxy}
\begin{center}
\begin{tabular}{c c c c}
\tableline
\tableline
Night  &  Filter  &  Magnitude  & Error \\
\tableline
2  &  U  &  19.57  &  0.45 \\
1  &  B  &  18.83  &  0.09 \\
1  &  V  &  17.53  &  0.04 \\
2  &  R  &  17.19  &  0.07 \\
2  &  I  &  16.47  &  0.07 \\
\tableline
\end{tabular}
\end{center}
\label{table:magnitudes} 
\end{table}

\clearpage

\begin{table}
\caption[Emission Line Data for NVSS 2146+82]{Emission Line Data for NVSS 2146+82}
\begin{center}
\begin{tabular}{l c c c c c c}
\tableline
\tableline
Species & $\lambda_{\rm red}$ & $z_{\rm red}$ & $\lambda_{\rm blue}$ &  $z_{\rm bl
ue}$ & 
Flux\tablenotemark{a} & Luminosity\tablenotemark{a} \\
   &  \AA     &   & \AA &  & $10^{-15}$ erg/sec/cm$^{2}$ & $10^{41}$ erg/sec $h_{50}^{-2}$ \\
\hline
$[$\ion{O}{2}$]$  $\lambda3727$ &  4262.5 & 0.1436 & 4269.6  & 0.1455  & 
\phn7.4$\pm$0.5  & \phn7.6$\pm$0.5 \\
$[$\ion{Ne}{3}$]$ $\lambda3869$ &  4425.3 & 0.1439 & 4432.6  & 0.1458  & 
\phn2.8$\pm$0.3  & \phn2.9$\pm$0.3 \\
$[$\ion{Ne}{3}$]$ $\lambda3967$\tablenotemark{b} &  &  & 4545.1  & 0.1456 & 
\phn0.9$\pm$0.1 & \phn0.9$\pm$0.1\\
H$\delta$                       &  4693.7 & 0.1443 & 4698.2  & 0.1454  & 
\phn0.3$\pm$0.1 & \phn0.3$\pm$0.1 \\
H$\gamma$                       &  4963.9 & 0.1436 & 4971.9  & 0.1455  & 
\phn1.4$\pm$0.2 & \phn1.4$\pm$0.2 \\
$[$\ion{O}{3}$]$ $\lambda4363$\tablenotemark{b}  &  &  & 4998.8  & 0.1457 & 
\phn0.8$\pm$0.1 & \phn0.8$\pm$0.1 \\
\ion{He}{2} $\lambda4686$       &  5360.8 & 0.1440 & 5367.5  & 0.1454  & 
\phn0.8$\pm$0.2 & \phn0.8$\pm$0.2  \\
H$\beta$                        &  5560.5 & 0.1438 & 5569.2  & 0.1456  & 
\phn3.0$\pm$0.4 & \phn3.1$\pm$0.4 \\
$[$\ion{O}{3}$]$ $\lambda4959$  &  5672.4 & 0.1439 & 5681.6  & 0.1457  & 
12.3$\pm$1.1 & 12.7$\pm$1.1 \\
$[$\ion{O}{3}$]$ $\lambda5007$  &  5727.2 & 0.1439 & 5736.4  & 0.1457  & 
35.7$\pm$3.2 & 36.9$\pm$3.3 \\
$[$\ion{O}{1}$]$ $\lambda6300$  &  7206.2 & 0.1438 & 7217.4  & 0.1456  & 
\phn1.6$\pm$0.4 & \phn1.7$\pm$0.4 \\
$[$\ion{N}{2}$]$ $\lambda6548$  &  7493.7 & 0.1444 & 7505.4  & 0.1462  & 
\phn1.7$\pm$0.3 & \phn1.8$\pm$0.3 \\
H$\alpha$                       &  7508.9 & 0.1442 & 7520.3  & 0.1459  & 
\phn8.9$\pm$1.0 & \phn9.2$\pm$1.0 \\
$[$\ion{N}{2}$]$ $\lambda6584$  &  7530.6 & 0.1438 & 7542.8  & 0.1457  & 
\phn5.1$\pm$0.6 & \phn5.3$\pm$0.6 \\
\tableline
\end{tabular}
\end{center}

\tablenotetext{a}{These values have been dereddened using a value
of $A_{V} = 0.9$.  Errors include only
calibration and measurement error, error in reddening is not included.}
\tablenotetext{b}{These lines were not resolved into a blue and red component; the
 values 
listed in the table were determined by fitting the profile with a single gaussian.
}

\label{table:linedata} 
\end{table}

\clearpage

\begin{table}
\caption[Redshifts of Candidate Cluster Members in the Field of NVSS 2146+82]
{Redshifts of Candidate Cluster Members in the Field of NVSS 2146+82}
\begin{center}
\begin{tabular}{c c c c c c}
\tableline
\tableline
Galaxy ID  &  $\alpha_{2000.0}$ &  $\delta_{2000.0}$ &  z  &  q  &  $m_{v}$ \\
\tableline
 1 & 21:42:18.5 &  81:55:34  & 0.242 & 5 & 20.0 \\
 2 & 21:42:56.3 &  81:48:29  & 0.378 & 3 & 20.2 \\
 3 & 21:42:58.5 &  81:57:40  & 0.350 & 5 & 19.4 \\
 4 & 21:44:47.8 &  81:56:15  & 0.145 & 6 & 18.8 \\
 5 & 21:45:24.5 &  81:55:05  & 0.144 & 6 & 19.3 \\
 6 & 21:45:27.7 &  81:57:54  & 0.267 & 6 & 19.4 \\
 7 & 21:45:54.8 &  81:53:23  & 0.135 & 6 & 18.4 \\
 8 & 21:46:08.8 &  81:48:08  & 0.123 & 6 & 20.1 \\
 9 & 21:46:24.3 &  81:57:43  & 0.243 & 6 & 18.4 \\
10 & 21:46:48.6 &  82:01:46  & 0.183 & 2 & 19.8 \\
11 & 21:47:05.7 &  81:52:35  & 0.144 & 6 & 18.3 \\
12 & 21:47:07.2 &  81:55:36  & 0.145 & 6 & 18.4 \\
13 & 21:47:20.4 &  81:51:40  & 0.149 & 1 & 19.7 \\
14 & 21:47:24.7 &  81:50:40  & 0.173 & 6 & 20.0 \\
15 & 21:47:53.7 &  81:53:43  & 0.208 & 2 & 19.4 \\
16 & 21:48:19.9 &  82:01:03  & 0.143 & 1 & 17.7 \\
17 & 21:48:44.3 &  81:56:59  & 0.148 & 6 & 18.9 \\
\tableline
\end{tabular}
\end{center}
\label{table:redshifts} 
\end{table}

\clearpage

\begin{table}
\caption[Giant Radio Galaxies]{Giant Radio Galaxies}
\begin{center}
\begin{tabular}{c c c c c c}
\tableline
\tableline
IAU Name  &  Other Name  &  z  &  LAS &  log$P_{1.4}$ &  LLS \\
          &              &     & (arcsec) & ($h_{50}^{-2}$ W Hz$^{-1}$) & ($h_{50}^{-1}$ Mpc)\\
\tableline
 1003+351 & 3C\,236   &  0.0989  & 2478  & 26.37 & 6.04\\
{\bf 2146+82}  & {\bf NVSS 2146+82} & {\bf 0.1450} & {\bf 1175} & {\bf 25.69} & {\bf 3.91 }\\
 0821+695 & 8C 0821+695 & 0.5380 &  402  & 26.30 & 2.94 \\
 1637+826 & NGC\,6251 &  0.0230  & 4500  & 24.73 & 2.89 \\
 0319-454 &           &  0.0633  & 1644  & 25.83 & 2.72 \\
 1549+202 & 3C\,326   &  0.0885  & 1206  & 26.08 & 2.67 \\
 1358+305 & B2 1358+305 & 0.2060 &  612  & 25.93 & 2.64 \\
 1029+570 & HB\,13    &  0.0450  & 2100  & 24.57 & 2.54 \\
 0503-286 &           &  0.0380  & 2400  & 25.23 & 2.48 \\
 1452-517 & MRC 1452-517 & 0.08  & 1218  & 25.66 & 2.48 \\
 0114-476 & PKS 0114-476 & 0.1460 & 702  & 26.51 & 2.36 \\
 1127-130 & PKS 1127-130 & 0.6337 & 297  & 27.53 & 2.30 \\
 0707-359 & PKS 0707-359 & 0.2182 & 492  & 26.71 & 2.21 \\
 1910-800 &           &  0.3460  &  366  & 26.65 & 2.18 \\
 0745+560 & DA\,240   &  0.0350  & 2164  & 25.39 & 2.07 \\
 0313+683 & WENSS 0313+683 & 0.0902 & 894 & 25.64 & 2.01 \\
\tableline
\end{tabular}
\end{center}
\label{table:giants}

\end{table}


\clearpage

\begin{thebibliography}{}

\bibitem[Abell 1958]{abell} Abell, G. O. 1958, \apjs, 3, 211 

\bibitem[Allington-Smith et al.\ 1993]{aezo} Allington-Smith, J. R.,
Ellis, R. S., Zirbel, E. L., \& Oemler, A., 1993, \apj, 404, 521 

\bibitem[Berkhuijsen 1986]{Berkhuijsen}
Berkhuijsen, E. M. 1986, \aap, 166, 257

\bibitem[Cardelli et al.\ (1989)]{cardelli} Cardelli, J. A., Clayton, G. C. \& 
Mathis, J. S. 1989, \apj, 345, 245 

\bibitem[Condon et al.\ 1998]{Condon98}
Condon, J. J., Cotton, W. D., Greisen, E. W., Yin, Q. F., Perley, 
R. A., Taylor, G. B., \& Broderick, J. J., 1998, \aj, 115, 1693

\bibitem[Cotter, Rawlings, \& Saunders 1996]{cot96}
Cotter, G., Rawlings, S., \& Saunders, R. 1996, \mnras, 281, 1081

\bibitem[Ebeling et al.\ 1998]{Ebeling98} Ebeling, H., Edge, A. C.,
Bohringer, H.,  Allen, S. W., Crawford, C. S., Fabian, A. C., Voges,
W. \& Huchra, J. P. 1998, \mnras, 301, 881

\bibitem[Fabbiano et al.\ (1984)]{fabbiano84} Fabbiano, G., 
Trinchieri, G., Elvis, M., Miller, L. \& Longair, M. 1984, \apj, 277, 115 

\bibitem[Fanaroff \& Riley 1974]{fan74}
Fanaroff, B. L., \& Riley, J. M. 1974, \mnras, 167, 31

\bibitem[Green, Anderson \& Ward 1992]{Green92} Green, P. J., Anderson, S. F. \& 
Ward, M. J. 1992, \mnras, 254, 30

\bibitem[Halpern \& Eracleous 1994]{he94} Halpern, J. P., \& Eracleous, M. 1994,
\apj, 433, L17

\bibitem[Heckman et al.\ 1986]{heckman86} Heckman, T. M., Smith, E. P., Baum, S. A., 
van Breugel, W. J. M., Miley, G. K., Illingworth, G. D., Bothun, G. D., \&
Balick, B., 1986, \apj, 311, 526

\bibitem[Hill \& Lilly 1991]{hill91} Hill, G. J., \& Lilly, S. J., 1991, \apj
367, 1

\bibitem[Kinney et al.\ 1996]{kinney96} Kinney, A. L., Calzetti, 
D. , Bohlin, R. C., McQuade, K. , Storchi-Bergmann, T.  \& Schmitt, H. R. 
1996, \apj, 467, 38 

\bibitem[Lacy et al.\ 1993]{lacy93} Lacy, M., Rawlings, S., Saunders, R.,
\& Warner, P. J. 1993, \mnras, 264, 721

\bibitem[Landolt (1992)]{landolt} Landolt, A. U., 1992, \aj, 104, 340


\bibitem[Lasker et al.\ 1990]{dss} Lasker, B. M., Sturch, C. R., McLean, B. J., 
Russell, J. L., Jenkner, H., \& Shara, M. M., 1990, \aj, 99, 2019

\bibitem[Longair \& Seldner 1979]{longair79} Longair, M. S. \& 
Seldner, M. 1979, \mnras, 189, 433 

\bibitem[Miller et al.\ (1999)]{miller99} Miller, N. A., Owen, F. N., Burns, J. O., 
Ledlow, M. J., \& Voges, W. 1999, preprint (astro-ph/9908244)

\bibitem[Monet et al.\ 1996]{monet96} Monet, D., Bird, A., Canzian, B., Harris, H., 
Reid, N., Rhodes, A., Sell, S., Ables, H., Dahn, C., Guetter, H., Henden,
A., Leggett, S., Levison, H., Luginbuhl, C., Martini, J., Monet, A., Pier, J., 
Riepe, B., Stone, R., Vrba, F., Walker, R.
1996, USNO-SA1.0, (Washington D.C.: U.S. Naval Observatory)

\bibitem[Mukai 1993]{Mukai93} Mukai, K. 1993, Legacy 3, 21-31

\bibitem[Munn et al.\ (1997)]{munn97} Munn, J. A., Koo, D. C., Kron, R. G., Majewski,
S. R., Bershady, M. A., \& Smetanka, J. J., 1997, \apjs, 109, 45

\bibitem[Nilsson et al.\ (1993)]{nilsson}Nilsson, K., Valtonen, M. J., Kotilainen, J., 
\& Jaakkola, T., 1993, \apj, 413, 453

\bibitem[Osterbrock 1989]{osterbrock} Osterbrock, D. E., 1989, Astrophysics of Gaseous
Nebulae and Active Galactic Nuclei, (Mill Valley: University Science Books)

\bibitem[Parma et al.\ 1996]{par96}
Parma, P., de Ruiter, H., Mack, K.-H., van Breugel, W., Dey, A., Fanti, R.,
\& Klein, U., 1996, \aap, 311, 49

\bibitem[Plucinsky et al.\ 1993]{Plucinsky93} Plucinsky, P. P., Snowden, S. L., 
Briel, U. G., Hasinger, G., \& Pfeffermann, E. 1993, \apj, 418, 519

\bibitem[Prestage \& Peacock 1988]{prestage88} Prestage, R. M. \& 
Peacock, J. A. 1988, \mnras, 230, 131

\bibitem[Rawlings \& Saunders 1991]{raw91} Rawlings, S., \& Saunders, R., 1991,
Nature, 349, 138

\bibitem[Rengelink et al.\ 1997]{WENSS}
Rengelink, R. B., Tang, Y., de Bruyn, A. G., Miley, G. K., Bremer, M.
N., R\"ottgering, H. J. A., Bremer, M. A. R. 1997, \aaps, 124,259

\bibitem[Sandage (1972)]{sandage72} Sandage, A. R. 1972, \apj, 178, 25

\bibitem[Saripalli et al.\ 1986]{sar86} Saripalli, L., Gopal-Krishna, Reich, W., 
\& K\"{u}hr, H. 1986, \aap, 170, 20

\bibitem[Schlegel et al.\ (1998)]{schlegel} Schlegel, 
D. J., Finkbeiner, D. P. \& Davis, M.  1998, \apj, 500, 525 

\bibitem[Simard-Normandin, Kronberg, \& Button (1981)]{Simon-Normandin}
Simard-Normandin, M., Kronberg, P. P., \& Button, S. 1981, \apjs, 45, 97

\bibitem[Snowden 1998]{Snowden98} 
Snowden, S. L. 1998, \apjs, 117, 233

\bibitem[Stark et al.\ 1992]{stark} Stark, A. A., Gammie, C. F., 
Wilson, R. W., Bally, J., Linke, R. A., Heiles, C., \& Hurwitz, M. 1992, \apjs, 79, 77

\bibitem[Stetson 1987]{stetson87} Stetson, P. B. 1987, \pasp, 99, 191

\bibitem[Strom \& Willis 1980]{str80}
Strom, R. G. \& Willis, A. G. 1980, \aap, 85, 36

\bibitem[Subrahmanyan \& Saripalli 1993]{sub93}
Subrahmanyan, R. \& Saripalli, L. 1993, \mnras, 260, 908

\bibitem[Tadhunter et al.\ 1998]{tad98} Tadhunter, C. N., Morganti, R., Robinson, A.,
Dickson, R., Villar-Martin, M., \& Fosbury, R. A. E. 1998, \mnras, 298, 1035

\bibitem[Taylor, Dyson, \& Axon 1992]{tda92} Taylor, D., Dyson, J. E., \& Axon,
D. J. 1992, \mnras, 255, 351

\bibitem[Valdes 1982]{valdes82} Valdes, F. 1982, FOCAS User's Manual, (Tucson:  NOAO)

\bibitem[Wan \& Daly (1996)]{wan96} Wan, L., \& Daly, R. A. 1996, \apj, 467, 145

\bibitem[Whittle 1989]{dmw89} Whittle, M. 1989, in Extranuclear Activity in Galaxies,
ed. E. Meurs \& R. Fosbury (Munich: ESO), 199

\bibitem[Willis, Strom, \& Wilson 1974]{wil74}
Willis, A. G., Strom, R. G., \& Wilson, A. S. 1974, {\it Nature}, 250, 625

\bibitem[Zirbel 1996]{zirbel96} Zirbel, E. L. 1996, \apj, 473, 713

\bibitem[Zirbel 1997]{zirbel97} Zirbel, E. L. 1997, \apj, 476, 489

\bibitem[Zirbel \& Baum 1995]{zirbel95} Zirbel, E. L., \& Baum, S. A. 1995, \apj,
448, 521

\bibitem[Zwicky et al.\ 1961]{zwicky} Zwicky, F., Herzog, E., Wild, P.,
Karpowicz, M., \& Kowal, C. 1961-68, Catalogue of Galaxies and Clusters
of Galaxies, (Pasadena: CIT)

\end{thebibliography}
\end{document}